\documentclass{article}

     \usepackage[preprint]{neurips_2024}

\usepackage[utf8]{inputenc} 
\usepackage[T1]{fontenc}    
\usepackage{hyperref}       
\usepackage{url}            
\usepackage{booktabs}       
\usepackage{amsfonts}       
\usepackage{nicefrac}       
\usepackage{microtype}      
\usepackage{xcolor}         

\usepackage{tabularx}
\usepackage{graphicx}
\usepackage{float}
\usepackage{subcaption}
\usepackage{bm}
\usepackage{makecell}
\usepackage{multirow}
\usepackage{caption}

\title{CrypticBio: A Large Multimodal Dataset for Visually Confusing Biodiversity}

\author{%
Georgiana Manolache\textsuperscript{1},   Gerard Schouten\textsuperscript{1}, Joaquin Vanschoren\textsuperscript{2} \\\\
  \textsuperscript{1}Fontys University of Applied Sciences, Eindhoven, The Netherlands\\
  \textsuperscript{2}Technical University of Eindhoven, Eindhoven, The Netherlands\\
  Correspondence: \texttt{g.manolache@fontys.nl} 
}

\begin{document}

\maketitle

\begin{abstract}
We present \textsc{CrypticBio}, the largest publicly available multimodal dataset of visually confusing species, specifically curated to support the development of AI models in the context of biodiversity applications.
Visually confusing or cryptic species are groups of two or more taxa that are nearly indistinguishable based on visual characteristics alone.
While much existing work addresses taxonomic identification in a broad sense, datasets that directly address the morphological confusion of cryptic species are small, manually curated, and target only a single taxon. 
Thus, the challenge of identifying such subtle differences in a wide range of taxa remains unaddressed.
Curated from real-world trends in species misidentification among community annotators of iNaturalist, \textsc{CrypticBio} contains 52K unique cryptic groups spanning 67K species, represented in 166 million images. 
Rich research-grade image annotations—including scientific, multicultural, and multilingual species terminology, hierarchical taxonomy, spatiotemporal context, and associated cryptic groups—address multimodal AI in biodiversity research.
For easy dataset curation,
we provide an open-source pipeline
\textsc{CrypticBio-Curate}.  
The multimodal nature of the dataset beyond vision-language arises from the integration of geographical and temporal data as complementary cues to identifying cryptic species.
To highlight the importance of the dataset, we benchmark a suite of state-of-the-art foundation models across \textsc{CrypticBio} subsets of common, unseen,  endangered, and invasive species, and demonstrate the substantial impact of geographical context on vision-language zero-shot learning for cryptic species.
By introducing \textsc{CrypticBio}, we aim to catalyze progress toward real-world-ready biodiversity AI models capable of handling the nuanced challenges of species ambiguity.
The data and the code are publicly available in the \href{https://georgianagmanolache.github.io/crypticbio/}{\textcolor{blue}{project website}}.
\end{abstract}

\section{Introduction}
\label{sections:introduction}
\begin{figure}[t]
    \centering
\includegraphics[width=\textwidth]{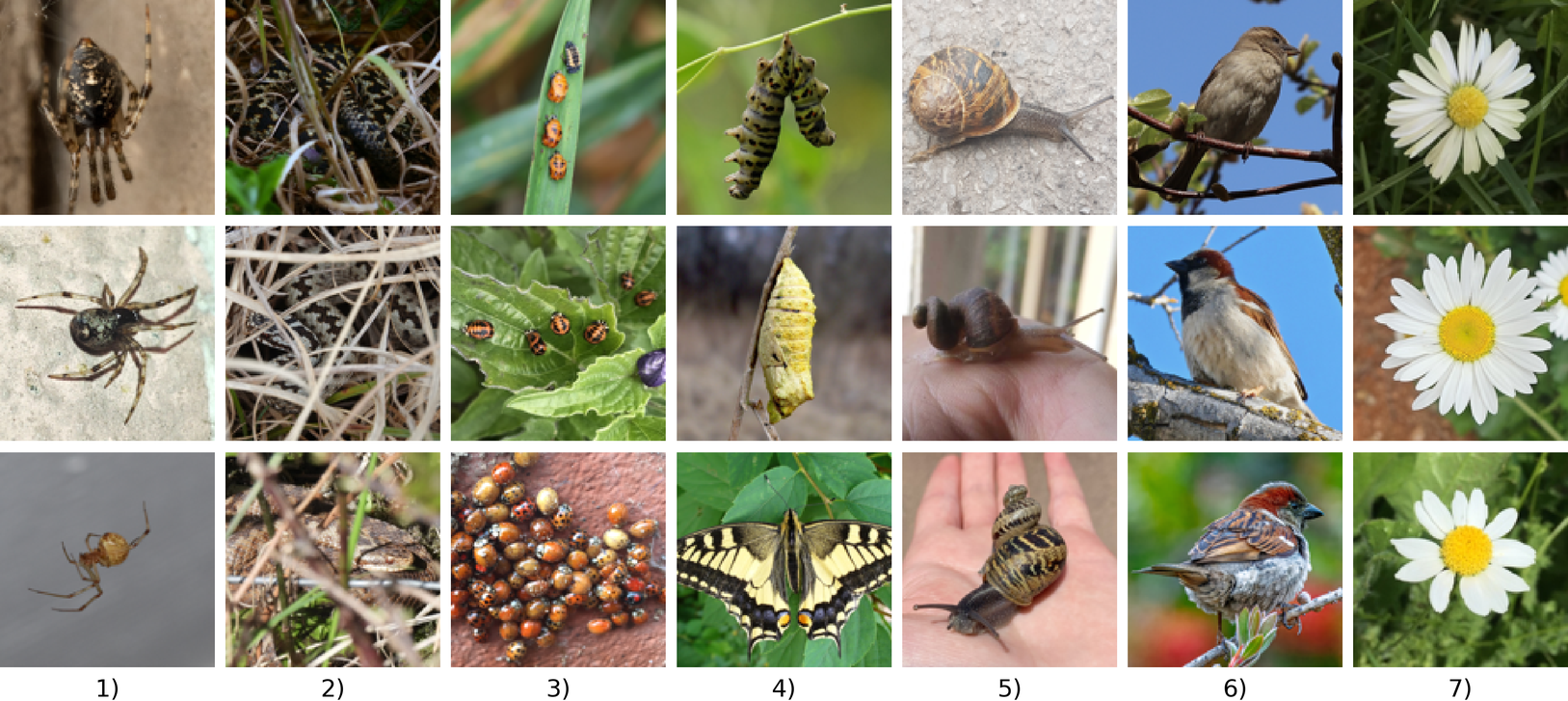}
    \caption{Challenges of biodiversity: (1) viewpoint variations (\textit{Parasteatoda tepidariorum}); (2) occlusion by other objects (\textit{Vipera berus}); (3) clutter (\textit{Harmonia axyridis}); (4) multiple life cycle stages (\textit{Papilio machaon}); (5) deformations (\textit{Cornu aspersum}); (6) intra-class variation (\textit{Passer domesticus}); (7) inter-class similarity (\textit{Bellis perennis}, \textit{Leucanthemum vulgare}, \textit{Chamomile matricaria}).}
    \label{fig:challenges}
\end{figure}

Advancements in AI are set to play a pivotal role in biodiversity conservation and ecological management as data in open citizen science platforms amasses.
iNaturalist~\cite{inaturalist2025} and Observation.org~\cite{observation2025} are well established citizen science platforms collecting biodiversity data worldwide, featuring annotated in-situ images of a wide range of species as well as contextual metadata such as geographical location and observation date.
AI-ready biodiversity datasets are a crucial part of the development, evaluation, and eventual deployment of machine learning (ML) systems, and many works already demonstrate their potential in species identification  \cite{stevens2024bioclip,yang2024biotrove,sastry2024taxabind}. 
As it turns out, however, species identification combines a unique set of challenges.
As illustrated in Figure~\ref{fig:challenges}, these include: (1) viewpoint variations; (2) occlusion by other objects; (3) clutter; (4) multiple life cycle stages; (5) deformations; (6) intra-class variation; (7) inter-class similarity. 
The latter two challenges are particularly hard: some species may have significant visual differences, while at the same time visual similarities in shape and color may exist between some species belonging to different classes. 
This morphological confusion makes it difficult even for humans to distinguish the species without deeper expertise, and subsequently limits the construction of trustworthy AI for biodiversity~\cite{hending2025cryptic}. 

Existing state-of-the-art datasets of text-annotated biodiversity images primarily curated from iNaturalist focus on taxa identification holistically. Notable examples include
\textsc{TreeOfLife-10M}~\citep{stevens2024bioclip} with over 10 million observations spanning 451K species, and \textsc{BioTrove}~\citep{yang2024biotrove} with over 161 million observations spanning 366K species, respectively. More recently, the multimodal dataset \textsc{TaxaBind-8K}~\citep{sastry2024taxabind} extending over 8K text-annotated biodiversity images with other contextual metadata like geographical location, environmental features, audio recordings, and satellite imagery, shows significant improvements on 2K bird species identification, using vision as the binding modality in a unified embedding space. 
Datasets that directly address the morphological confusion of groups of two or more species are significantly smaller, manually curated, and focused on a single taxon~\citep{chiranjeevi2023deep,pinho2022squamata,cao2024bats,kim2025parrots}.
Thus, the challenge of identifying subtle differences in a wide range of taxa remains to be addressed.

In this paper, we challenge the biodiversity AI research by curating and releasing \textbf{\textsc{CrypticBio}}, a \textbf{multimodal dataset} comprising over \textbf{166 million images} of \textbf{52K unique visually confusing species groups} spanning \textbf{67K species}. 
Table \ref{table:comparisons} summarizes how \textsc{CrypticBio} compares to prior biodiversity datasets in scale and annotation richness, and Table \ref{table:existing_benchmarks} contrasts it with existing cryptic species benchmarks (which are several orders of magnitude smaller).
As this work is intended to have a direct impact on the use of AI for biodiversity research, we hope it will provide valuable insights to researchers seeking to better understand biodiversity.
Our main contributions include:
(1) the largest multimodal cryptic species dataset to date; 
(2) broad taxonomic coverage beyond prior single-taxon studies, 
(3) enriched metadata (locations, dates, multicultural and multilingual names) enabling new multimodal research; 
(4) an open-source pipeline for easy dataset curation (\textsc{CrypticBio-Curate}); 
(5) evaluation benchmarks demonstrating the utility of context in species identification.

The remainder of the paper introduces the \textsc{CrypticBio} dataset and its relation to existing work (Section \ref{sections:dataset}), outlines the curation methodology (Section \ref{sections:data_curation}), presents benchmark tasks and zero-shot results using CLIP-style models (Section \ref{sections:models_benchmakrs}), and concludes with a summary and discussion of limitations (Section \ref{sections:conclusion}). Supplementary material includes implementation details, extended results, and dataset access instructions.

\begin{figure}[t]
    \centering
    \includegraphics[width=\linewidth]{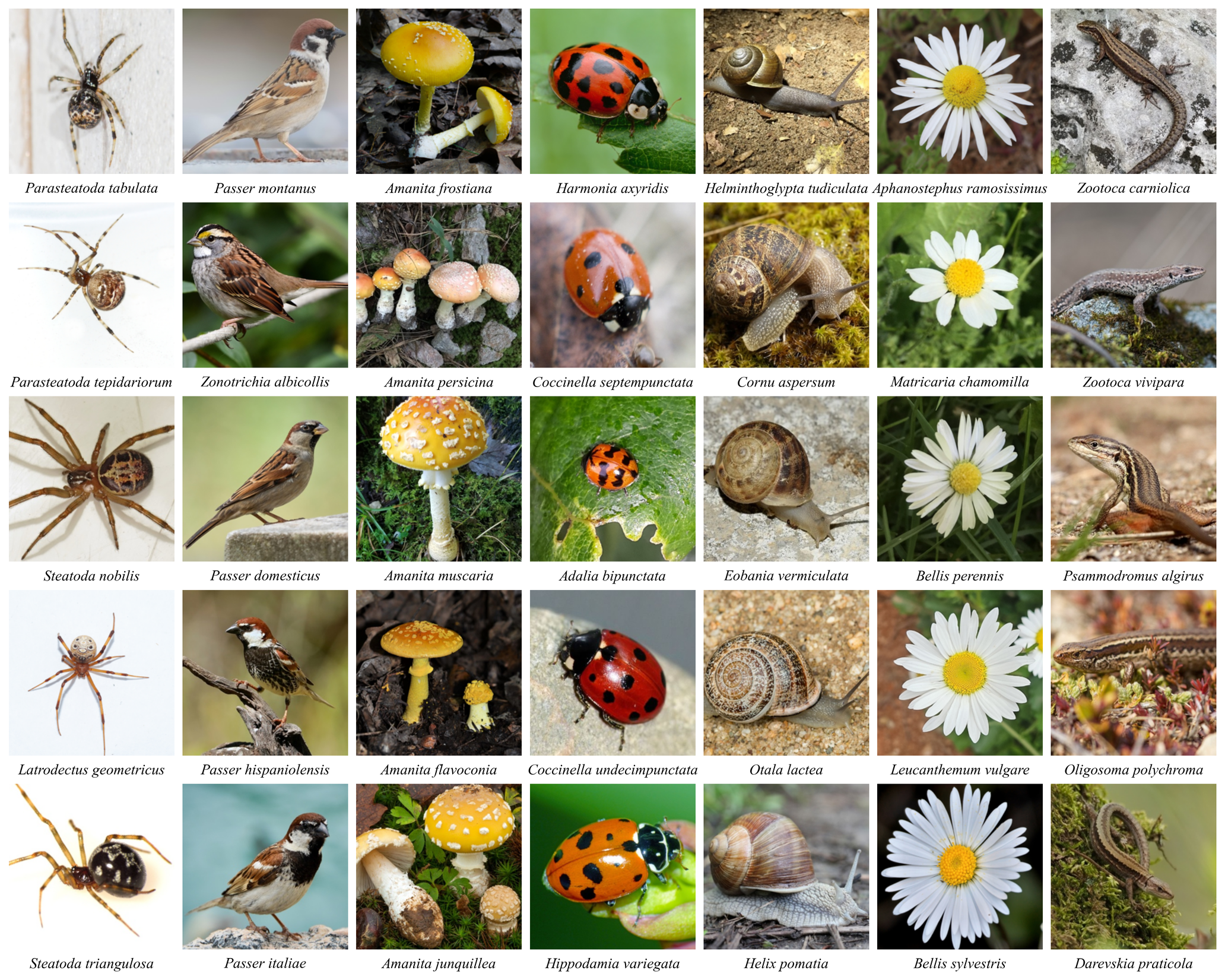} 
    \caption{  
    Example of cryptic species in \textsc{CrypticBio}. Each column shows from left to right cryptic groups from \textit{Arachnida}, \textit{Aves}, \textit{Insecta}, \textit{Plantae}, \textit{Fungi}, \textit{Mollusca}, and \textit{Reptilia}, taxa representative in biodiversity conservation and policy change supervision.}
    \label{fig:sample}
\end{figure}

\section{\textsc{CrypticBio} Dataset}
\label{sections:dataset}

\begin{figure}[t]
\centering
\includegraphics[width=\linewidth]
{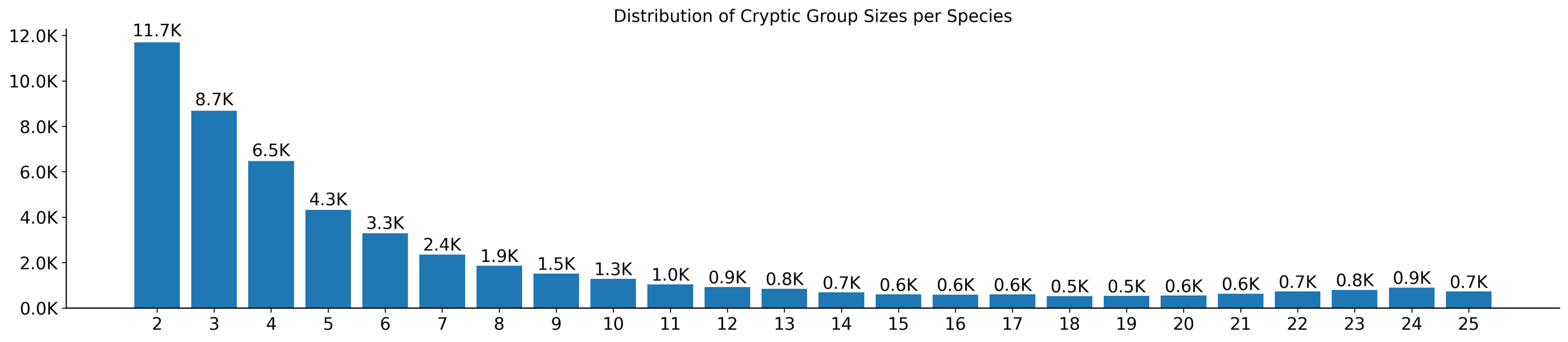}
        \label{fig:cryptic_group_distribution}
   \caption{Cryptic group size distribution in \textsc{CrypticBio}. The long-tailed distribution suggests that the majority are divided into a small number of cryptic entities.}
    \label{fig:cryptic_distributions}
\end{figure}

\textsc{CrypticBio} comprises over \textbf{166 million images} from \textbf{52K unique cryptic groups} spanning \textbf{67K species}. 
Figure~\ref{fig:sample} showcases cryptic species group examples over the representative taxa in biodiversity conservation and policy change supervision, while Figure \ref{fig:cryptic_distributions} details the cryptic species group size distribution (see details in supplementary material \ref{supplementary_material:ethics_statement} and \ref{supplementary_material:dataset_details}). 
The dataset is curated from \textbf{research-grade citizen science observations} provided by the Global Biodiversity Information Facility (GBIF)~\cite{gbif2025}, containing validated data from iNaturalist, a source of demonstrated 95\% annotation reliability~\cite{inaturalist2024accuracy}, and Observation.org, an expert exclusive data validation source.
In iNaturalist, observations are graded research quality at the genus, species, or subspecies under the following criteria: (1) at least two identifications (including the observer's); and (2) there is an agreement among at least two-thirds of identifiers~\cite{inaturalist2025quality}. 
Observation.org data is validated through: (1) a structured review process in which only expert validators assess the presence and quality of supporting multimedia; or (2) a computer vision system, available through the Nature Identification API (NIA)~\cite{observation2025nia}, validates with high confidence~\cite{observation2025validation}.

\begin{table}[t]
\caption{\textsc{CrypticBio} comparable datasets.}
\label{table:comparisons}
\centering
{\scriptsize 
\begin{tabular}{ c  c  c  c  c  c } 
\toprule
\textbf{Datasets} &  \textbf{Images}& \textbf{Species} & \textbf{Annotations} & \textbf{Source} & \textbf{Features}\\
\midrule
\makecell{\textbf{\textsc{CrypticBio}}} & 166.5M & 67.1K & \makecell{ (multicultural and   \\ multilingual) vernacular \\scientific terms, 
taxonomic \\ hierarchy, location, date, \\ cryptic species group} & \makecell{ GBIF (iNaturalist and \\Observation.org), \\GBIF Backbone \\ Taxonomy~\cite{gbif2023backbone}, \\ iNaturalist  \\ Taxonomy~\cite{inaturalist2025taxonomy} } & \makecell{multimodal, \\ cryptic \\ species group \\ (52.7K groups)}\\
\midrule
\makecell{\textbf{\textsc{BioTrove}}~\cite{yang2024biotrove}}  & 161.9M  & 366.6K & \makecell{vernacular, scientific terms,\\ taxonomic hierarchy}& iNaturalist & \makecell{biased vernacular \\ species terminology  }\\
\midrule
\makecell{\textbf{\textsc{TreeOfLife-10M}}~\cite{stevens2024bioclip}}  & 10.4M  & 454.1K & \makecell{vernacular, scientific terms,\\ taxonomic hierarchy} & \makecell{ iNaturalist,\\ Encyclopedia \\of Life (EOL)\cite{eol2025},\\ \textsc{BioScan-1M}~\cite{gharaee2023step}} &  \makecell{biased vernacular \\ species terminology }\\
\midrule
\makecell{\textbf{\textsc{TaxaBind-8K}}~\cite{sastry2024taxabind}}  & 8.8K  & 2.2K & \makecell{vernacular, scientific terms, \\taxonomic hierarchy, location,\\ environmental features, audio\\  recording, satellite imagery} & \makecell{iNaturalist, \\ iNat2021\cite{van2021benchmarking}, \\ Santinel-2\cite{s2maps2023}, \\ WorldClim-2.1\cite{fick2017worldclim} }& \makecell{multimodal,\\ bird species\\ exclusive}\\
\bottomrule
\end{tabular}
}
\end{table}

\begin{table}[t]
\caption{Existing benchmarks; each represent one cryptic group. Our new benchmarks are described in Section \ref{section:benchmarks}.}
\label{table:existing_benchmarks}
\centering
{\scriptsize 
\begin{tabular}{ c  c  c  c  c  c } 
\toprule
\textbf{Taxon} & \textbf{Benchmark} & \textbf{Images}& \textbf{Species} & \textbf{Annotations} & \textbf{Source} \\
\midrule
\textit{Aves} & \makecell{\textbf{\textsc{Amazon Parrots}}~\cite{kim2025parrots}} & 14K & 35 & \makecell{scientific terms} & \makecell{iNaturalist, eBird~\cite{ebird2021}, \\Google Images} \\
\midrule
\multirow{4}{*}{
\textit{Insecta}} & \makecell{\textbf{\textsc{Bumble Bees}}~\cite{spiesman2021assessing}\\ (not publicly available)}  & 89K & 36 & \makecell{scientific terms} & \makecell{iNaturalist, Bumble Bee \\Watch~\cite{hatfield2024bumblebeewatch}, BugGuide~\cite{bugguide2025}} \\
\cmidrule(r){2-6}
& \makecell{\textbf{\textsc{Confounding Species}}~\cite{chiranjeevi2023deep}\\ (not publicly available)}  & 100 & 10 & \makecell{scientific term}& iNaturalist \\
\midrule
\textit{Mammalia} & \makecell{\textbf{\textsc{Chiroptera Rhinolophidae}}\\ \textbf{\textsc{Rhinolophus}}~\cite{cao2024bats}}  & 293 & 7 & \makecell{scientific terms} & \makecell{personal collection \\ during field surveys} \\
\midrule
\multirow{3}{*}{\textit{Reptilia}} & \makecell{\textbf{\textsc{Sea Turtles}}~\cite{baek2024turtles} \\(not publicly available)}  & 6.9K & 36 & \makecell{vernacular, \\scientific terms} & \makecell{Internet} \\
\cmidrule(r){2-6}
& \makecell{\textbf{\textsc{Squamata Lacertidae}}\\ \textbf{\textsc{Podarcis}}~\cite{pinho2022squamata}}  & 4.0K & 9 & \makecell{scientific terms} & \makecell{personal collection \\ during field surveys} \\
\bottomrule
\end{tabular}
}
\end{table}

Images in \textsc{CrypticBio} are annotated with detailed taxonomic descriptions and observation context, enabling extensive filtering and analysis. 
As summarized in Table~\ref{table:annotations}, every observation at species level includes its scientific name and associated \textbf{six-tier taxonomic hierarchy} deterministically derived from GBIF Backbone Taxonomy~\cite{gbif2023backbone}.

While the inclusion of common or vernacular terminology alongside Latin binomials has been recognized as an important step toward accessibility and inclusivity in biodiversity datasets~\cite{stevens2024bioclip}, relying solely on English risks marginalizing indigenous and culturally specific naming traditions. Moreover, even within the English-speaking world, regional naming conventions can introduce their own biases. For example, the species \textit{Perisoreus canadensis} is commonly known as the \textit{Canada Jay} in Canada, yet referred to as the \textit{Gray Jay} in the United States~\cite{luccioni2023bugs}. Despite this variability, existing biodiversity datasets standardize to a single vernacular name per species, overlooking the cultural and linguistic diversity embedded in naming practices.
We believe integrating \textbf{multicultural and multilingual species vernacular names} preserves ecological knowledge and equity, and increases inclusivity and cultural reach. 
Therefore, we enrich species scientific names with vernacular naming practices as listed in the iNaturalist Taxonomy Archive~\cite{inaturalist2025taxonomy}.

\begin{figure}[t]
    \centering
    \begin{subfigure}{\textwidth}
     \centering
\includegraphics[width=\linewidth]{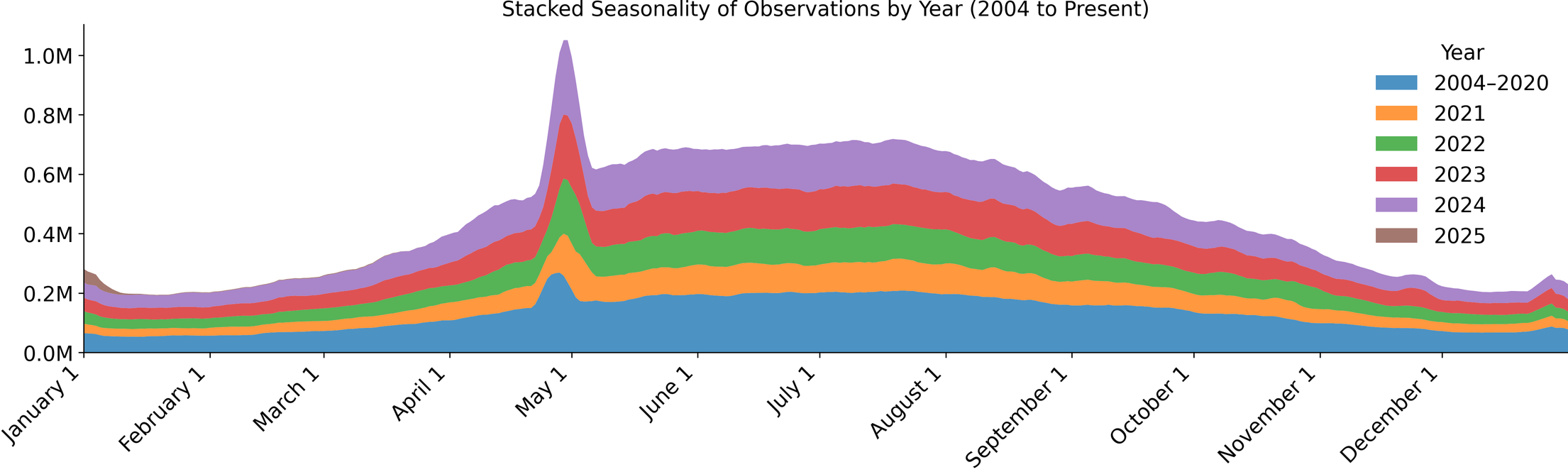}
        \label{fig:date}
    \end{subfigure}
    \begin{subfigure}{\textwidth}
    \includegraphics[width=\linewidth]{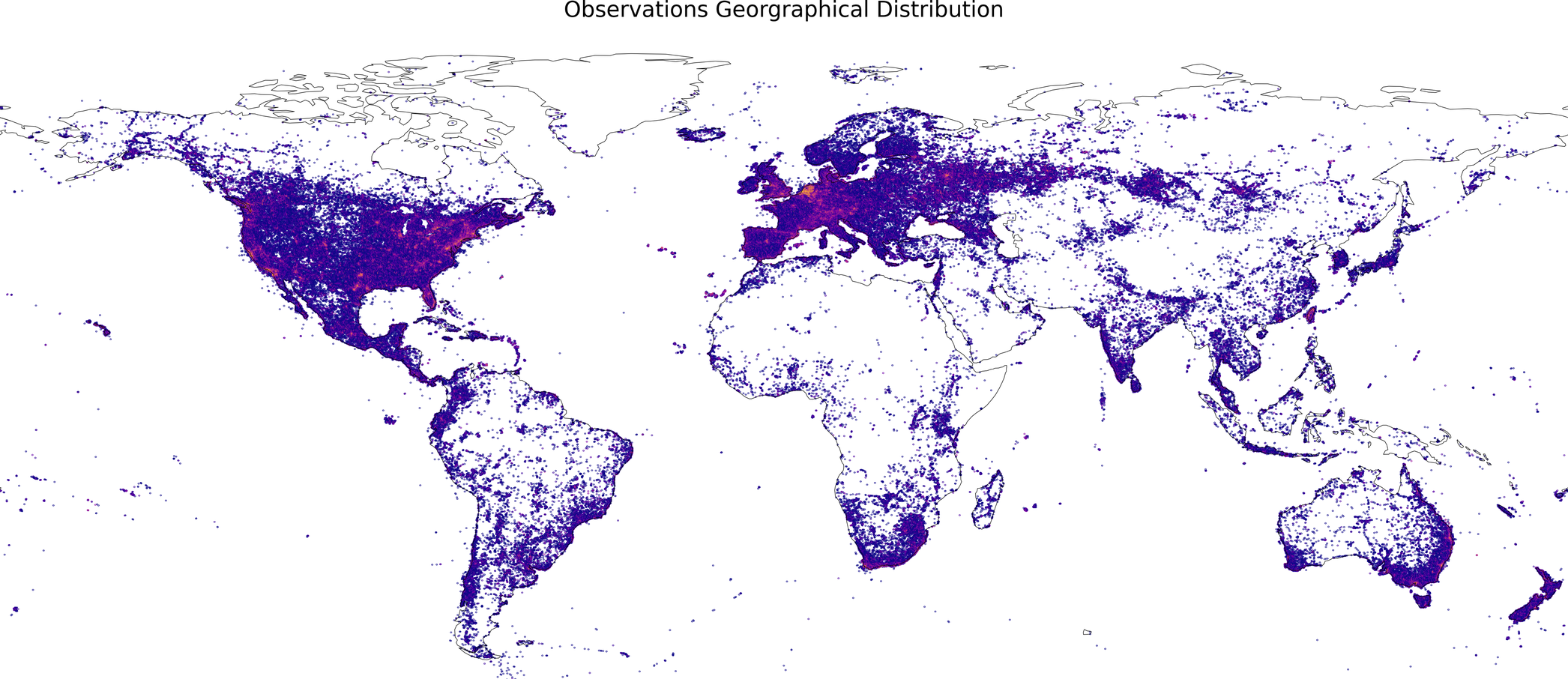}
        \label{fig:map}
    \end{subfigure}
    \caption{Spatiotemporal distribution of \textsc{CrypticBio}: 
    (top) stacked seasonality distribution; (bottom) geographical distribution. Majority of records are concentrated in Europe and North America, with a seasonal peak in observations during May.}
    \label{fig:distributions}
\end{figure}

\begin{figure}[t]
    \centering
    \includegraphics[width=\linewidth]{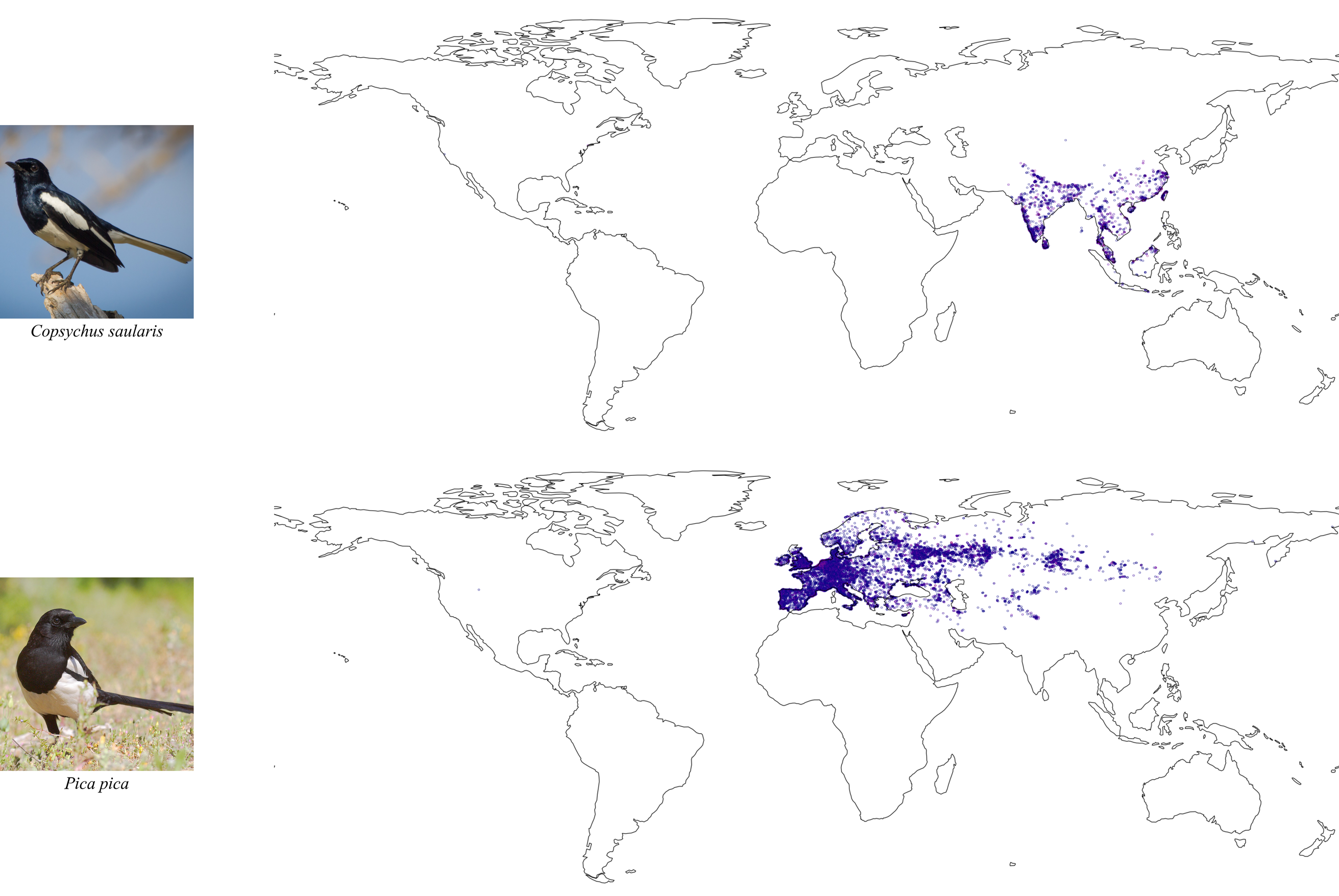} 
    \caption{The importance of geospatial information demonstrated
by two visually similar species and their distinct patterns in geospatial locations from \textsc{CrypticBio}.}
\label{fig:similar_georgraphical_distribution}
\end{figure}

Building on prior work demonstrating that spatiotemporal priors improve species identification~\cite{cole23,diao2022metaformer,sastry2023ld}, we integrate \textbf{spatiotemporal context as an additional modality} which can then eventually be aligned in a common embedding space. Figure \ref{fig:distributions} details the spatiotemporal distribution of the dataset.
Cryptic species have historically emerged as a consequence of biogeographic isolation (natural barriers, such as rivers, mountain ranges, or deserts; deforestation; agricultural expansion; or man-made structures) which disrupted gene flow between populations and ultimately promoted allopatric divergence over evolutionary timescales~\cite{hending2025cryptic}.
We hypothesize that the integration of temporal (date) and spatial (geographical coordinates) context provide complementary cues beyond visual appearance alone and ultimately enhance the identification of cryptic species. 
Figure \ref{fig:similar_georgraphical_distribution} illustrates an example of two cryptic bird species that have distinct geospatial distribution patterns and it is easy to tell them apart based on the location.

In our dataset, cryptic species groups are derived from \textbf{iNaturalist's records of historical misidentifications}. 
When an observation originally identified as species A is later reclassified as species B, species B is designated as a commonly misidentified counterpart of A. 
This data-driven cryptic species group curation reflects real-world trends in species misidentification among community annotators.
By leveraging this organically generated confusion structure, we design a benchmark that reflects the practical challenges of fine-grained species identification—particularly in cases where even skilled annotators struggle to distinguish between species.

Additionally, to support reproducibility and extensibility, we release the full data curation pipeline \textsc{CrypticBio-Curate}, enabling streamlined access, manipulation, and image download via raw URLs for \textsc{CrypticBio} subsets curation.

\textbf{License} Only images released under a Creative Commons (CC) license are included, ensuring that the dataset is openly available for public research and non-commercial use.

\textbf{Geoprivacy}
We include geolocation metadata for all records, relying on the source platforms’ automated and user-specified obscuration for sensitive species~\cite{observation2025geoprivacy,inaturalist2023geoprivacy}. This means endangered or protected taxa have deliberately imprecise coordinates, in line with geoprivacy best practices.

\textbf{Offensive content} 
We opted not to remove occasional graphic images (e.g. predation or roadkill) to maintain ecological authenticity. These instances are infrequent, but we advise users—especially when deploying models or visualizations—to be mindful that some images may be upsetting to general audiences.

\textbf{Responsible use} Models trained on this data should not be used for unlawful wildlife tracking or poaching; we provide the data to support conservation efforts and ecological research.

\textbf{Privacy} We strictly exclude all personally identifiable information (PII) from the metadata associated with the dataset, ensuring that fields such as observer names and email addresses are removed. 
However, we acknowledge that in rare cases, PII may still be visible within the image content itself; for example, faces of individuals, vehicle license plates, distinctive property features, or GPS location markers embedded in the media. 
While such occurrences are unintended and infrequent, users of the dataset should be aware of this residual risk when analyzing or displaying images.

\begin{table}[t]
\caption{\textsc{CrypticBio} annotations enumerate 15 fields provided in 627 Parquet formats; the dataset is openly available (for download and browsing) on \href{https://huggingface.co/datasets/gmanolache/CrypticBio}{\textcolor{blue}{HuggingFace Datasets}}.}
\label{table:annotations}
\centering
\small
\begin{tabularx}{\textwidth}{l X} 
\toprule
\textbf{Type} & \textbf{Description} \\
\midrule 
Species scientific name & Scientific species name (Latin binomial), represented as a string in field \texttt{scientificName}.\\
Species vernacular name(s) & Multicultural species common or vernacular names (i.e., \textit{Perisoreus canadensis} is commonly referred to as the \textit{Canada Jay} in Canada, while in the United States is referred to as \textit{Gray Jay}), represented in a sequence of strings in field \texttt{vernacularName}.\\
Taxonomic hierarchy & Species (primary) taxonomic hierarchy deterministically derived from species scientific name, represented as strings in separate fields: \texttt{kingdom}, \texttt{phylum}, \texttt{class}, \texttt{order}, \texttt{family}, \texttt{genus}.\\
Date & The date when the species was observed (separated DD, MM, YYYY), represented in fields \texttt{day}, \texttt{month}, \texttt{year}.\\
Geographical location & Latitude, longitude coordinates where the species was observed (decimals), represented in two fields \texttt{decimalLatitude} and \texttt{decimalLongitude}.\\
Cryptic species group & One or more species misidentified with the focal species, noted by scientific name, represented in a sequence of strings in field \texttt{crypticGroup}.\\
URL & Downloadable image link from Naturalist and Observation.org repositories, represented as a string in field \texttt{url}.\\ 
\bottomrule
\end{tabularx}
\end{table}

In comparison to prior biodiversity datasets (e.g., \textsc{TreeOfLife-10M}~\cite{stevens2024bioclip} and \textsc{BioTrove}~\cite{yang2024biotrove}) that take a broad but coarse approach, \textsc{CrypticBio} is the first to specifically target cryptic species at massive scale, with 166M images across 52K cryptic groups—significantly enriching each observation with spatiotemporal context and multicultural vernacular names. Compared to the only other multimodal dataset \textsc{TaxaBind-8K} (focused exclusively on 2K bird species), \textsc{CrypticBio} covers 67K species across diverse taxa with vision+language+geography modalities, making it a general, real-world AI-for-ecology benchmark.

\section{Data Curation}
\label{sections:data_curation}
\textbf{Cryptic species challenge}
While iNaturalist provides a "Similar Species" tab on individual taxon pages, which offers insights into species commonly confused by community annotators, it is not currently easily accessible.
The data is embedded in dynamic web elements and not easily exposed through the public iNaturalist API, nor is it available as part of their bulk data exports, adding complexity in data curation.

\textbf{Curation and filtering} 
To assemble a taxonomically diverse dataset utilizing GBIF's data portal, we apply a series of structured filters to occurrence records. GBIF comprises over 217M  occurrences as of 2025-04-13, originating from citizen science sources iNaturalist Research-Grade Observations and Observation.org, stored in a Darwin Core Archive standard vocabulary (CSV files \texttt{occurrence}, \texttt{multimedia} and additional metadata \texttt{eml.xml} and  \texttt{meta.xml}). 
We select observations of most frequent taxa from  
\textit{Animalia} (classes \textit{Arachnida}, \textit{Aves}, \textit{Insecta}, \textit{Mollusca}, \textit{Reptilia}), \textit{Plantae}, and \textit{Fungi} kingdoms, and join each \texttt{occurrence} and \texttt{multimedia} files, discarding irrelevant columns. 
We filter observations CC licensed image files (bird observations may include audio or video media files) made only species level and enrich them with primary taxonomic hierarchy levels (kingdom, phylum, class, order, family, and genus~\cite{iczn1999}), as well as multicultural English vernacular species terminology from iNaturalist Taxonomy~\cite{inaturalist2025taxonomy}. 
We retain the temporal (date) and spatial (latitude and longitude) metadata associated with each occurrence as additional contextual information. 
The associated images are referenced by downloadable URLs, ensuring direct access to the visual media for each observation (see detailed \textsc{CrypticBio} curation in supplementary material~\ref{supplementary_material:datasets_suite}).

To obtain species-level similarity data, we implement a targeted web scraping pipeline that systematically queries and parses the "Similar Species" section from individual taxon pages on iNaturalist.
We restrict our data collection to species-level matches that also appear in GBIF-filtered observations. 
Cryptic groups are structured using scientific names, based on curated information from iNaturalist, and stored in JSON format.
After extracting these groups, we merge them with GBIF observations, ensuring that we also retain species that do not have their own cryptic group but are listed as members of another species’ group.


As outlined, we release \textsc{CrypticBio-Curate}, a configurable preprocessing pipeline that streamlines the preparation of biodiversity datasets for multimodal learning. The pipeline (1) loads raw metadata (e.g., species scientific names, geolocations, image URLs); (2) applies customizable filters to ensure data quality (such as balancing class distributions); (3) downloads associated images; and (4) outputs a curated dataset in a standardized format.

\section{Models and Benchmarks}
\label{sections:models_benchmakrs}
\subsection{New benchmarks}
\label{section:benchmarks}

\begin{figure}[t]
    \centering
        \centering        \includegraphics[width=\linewidth]{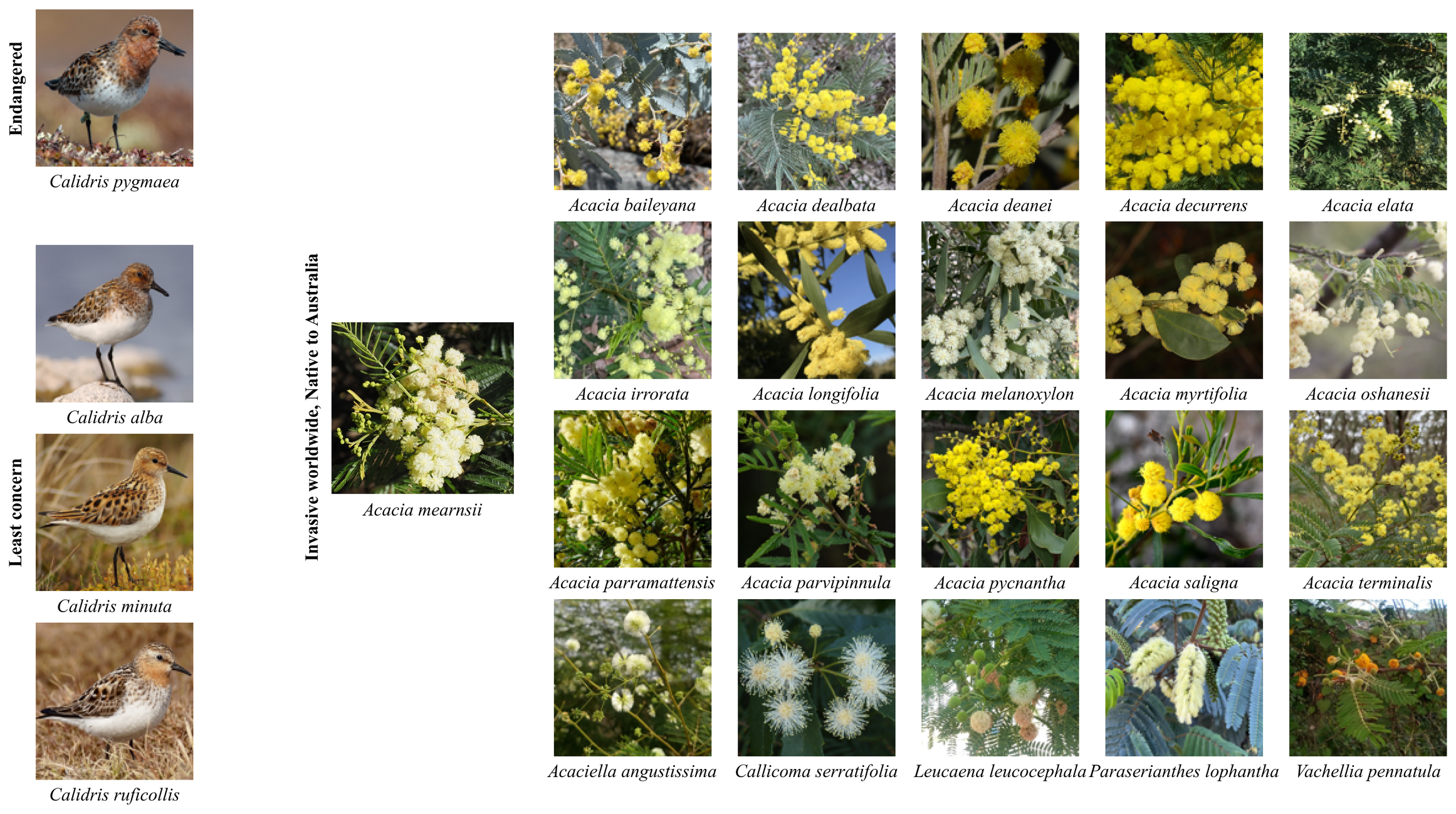}

    \caption{Example images from \textsc{CrypticBio} benchmarks: (left) \textsc{CrypticBio-Endangered} \textit{Calidris pygmaea} cryptic species group; (right) \textsc{CrypticBio-Invasive} \textit{Acacia mearnsii} cryptic species group.}
    \label{fig:sample_benchmarkss}
\end{figure}

We curate four new cryptic species benchmark datasets, complementary to existing benchmark listed in Table~\ref{table:existing_benchmarks}.  
Figure \ref{fig:sample_benchmarkss}  illustrates example images from our new benchmarks, which include both endangered and invasive cryptic species.
We rigorously balance species distribution to enables more reliable and equitable cryptic species group identification (see new benchmark details in supplementary material \ref{supplementary_material:new_benchmarks}): for all our benchmarks we randomly select 100 samples from each species in a cryptic group where there are more than 150 observation per species. 

\textbf{\textsc{CrypticBio-Common}} We curate one common species (species with >10K observations) from \textit{Arachnida}, \textit{Aves}, \textit{Insecta}, \textit{Plantae}, \textit{Fungi}, \textit{Mollusca}, and \textit{Reptilia} and associated cryptic group, spanning n=158 species. 

\textbf{\textsc{CrypticBio-CommonUnseen}} To assess performance on common species from \textsc{CrypticBio-Common} not encountered during model training, we specifically curate a subset containing data from 01-09-2024 to 01-04-2025, spanning n=133 species.  

\textbf{\textsc{CrypticBio-Endangered}} We propose a cryptic species subset of global IUCN Red List~\cite{iucn2024} endangered species. We select one endangered species from \textit{Arachnida}, \textit{Aves}, \textit{Insecta}, \textit{Plantae}, \textit{Fungi}, \textit{Mollusca}, and \textit{Reptilia} and associated cryptic group, spanning n=37 species.

\textbf{\textsc{CrypticBio-Invasive}}  We also propose a cryptic species subset of invasive alien species (IAS) according to global 
the Global Invasive Species Database (GISD)~\cite{gisd2025}. IAS are a significant concern for biodiversity as their records appear to be exponentially rising across the Earth, and their geographical context is crucial~\cite{mormul2022invasive}.
We select one invasive species from \textit{Aves}, 
\textit{Fungi}, \textit{Insecta}, and \textit{Plantae} and associated cryptic group, spanning n=72 species.

\subsection{Experiments}
\label{section:experiments}

\textbf{Models} 
We evaluate state-of-the-art CLIP-style models trained on biodiversity data using the scientific and vernacular terminology of species.  We use  \textsc{BioCLIP} \cite{stevens2024bioclip}; \textsc{BioTrove}'s \textsc{BioCLIP} ViT-B-16 and OpenAI ViT-B-16 fine-tuned variants \cite{yang2024biotrove}; and \textsc{TaxaBind} \cite{sastry2024taxabind} as image-only baseline models.
For multimodal learning, we add embeddings obtained from the image encoders to those obtained from \textsc{TaxaBind} location and environmental features encoders, which are then used for zero-shot classification. We collect from WorldClim-2.1\cite{fick2017worldclim} environmental features for each observation based on the location metadata,
which are then passed through \textsc{TaxaBind}'s environmental encoder.

\textbf{Metrics} We evaluate top-1 zero-shot accuracy across all new and existing benchmarks. 
We include a 95\% confidence intervals for all reported metrics, calculated using binomial proportion confidence interval method (denoted as $\pm$). 
Furthermore, we compute an aggregate performance metric, which represents the weighted average accuracy over all classes across the benchmarks.
To assess the significance of pairwise performance differences between models, we use McNemar’s test (p-value < 0.05).

\textbf{Results} Table~\ref{table:results} reports the performance on various benchmarks (see also extensive results in supplementary material \ref{supplementary_material:extended_results}). 
We find that location embeddings significantly improve model performance on zero-shot image classification for cryptic species (p-value < 0.05). 
While part of the observed gain from adding image and location embeddings likely reflects the real-world geographic separation of morphologically similar species, it is also possible that data collection is geographically biased, with certain species more frequently observed and labeled in particular regions. 
In such cases, location embeddings may act as proxies for latent biases in the training distribution, effectively anchoring predictions in more probable species given past observer behavior and sampling hotspots. However, in our setting, the evaluation datasets are randomly sampled from the full data distribution, without explicit regional or taxonomic filtering. 
This suggests that the performance boost is not merely an artifact of overfitting to spatial bias, but rather a reflection of how location embeddings can capture species ranges and statistical tendencies (i.e. regional observation frequencies) present in the broader data. 
We encourage the AI community to create new subsets of \textsc{CrypticBio} for various regions and measure performance against current benchmarks. 

\textbf{Limitations} 
We recognize that this data-driven identification of cryptic groups may miss rarely observed lookalike species. In future expansions, incorporating expert knowledge or targeted sampling of under-reported taxa could help capture cryptic relationships that have not yet appeared in crowd-sourced data.

Although our dataset includes temporal metadata, we have not systematically evaluated model performance when explicitly embedding this modality. This limits our understanding of how well the model leverages temporal patterns. While we still report significant zero-shot results using pretrained embeddings derived from different biodiversity datasets, few-shot learning could further enhance performance by enabling task-specific adaptation with minimal supervision.

\begin{table}[t]
\centering
\caption{Zero-shot learning on various models and benchmarks: I / L / E refers to image / location / environmental features embeddings; AP refers to \textsc{Amazon Parrots}~\cite{kim2025parrots} n=35   species; SLP refers to \textsc{Squamata Lacertidae Podarcis}~\cite{pinho2022squamata} n=9 species; CRR refers to \textsc{Chiroptera Rhinolophidae Rhinolophus}~\cite{cao2024bats} n=7 species; CB-C refers to \textsc{CrypticBio-Common} n=158 species; CB-CU refers to \textsc{CrypticBio-CommonUnseen} n=133; CB-E refers to \textsc{CrypticBio-Engendered} n=37 species; CB-I refers to \textsc{CrypticBio-Invasive} n=72 species; WA refers to weighted average; BC refers to \textsc{BioCLIP}; BT-B refers to \textsc{BioTrove-CLIP-BioCLIP}; BT-O refers to \textsc{BioTrove-CLIP-OpenAI}; TB refers to \textsc{TaxaBind}. We mix scientific and common terminology were avaiable.}
\label{table:results}
\centering
\scriptsize
{
\begin{tabularx}{\textwidth}{X c c c c c c c c c} 
\toprule
\textbf{Model} & \textbf{} & \textbf{AP} & \textbf{SLP} & \textbf{CRR} & \textbf{CB-C} & \textbf{CB-CU} & \textbf{CB-E} & \textbf{CB-I} & \textbf{WA} \\
\midrule
\textbf{BC} & I & 18.1 {\scriptsize ± 1.26} & 14.3 {\scriptsize ± 1.14} & \textbf{36.1 {\scriptsize ± 1.57}} & 42.7 {\scriptsize ± 1.61} & 45.7 {\scriptsize ± 1.63} &  51.1 {\scriptsize ± 1.63} & 49.1 {\scriptsize ± 1.63} & 44.36 \\
\textbf{BT-B} & I & 14.1 {\scriptsize ± 1.14} & 14.1 {\scriptsize ± 1.14} & 16.0 {\scriptsize ± 1.20} & 58.9 {\scriptsize ± 1.61} & 61.6 {\scriptsize ± 1.59} &  45.8 {\scriptsize ± 1.63} & 58.3 {\scriptsize ± 1.62} & 51.54 \\
\textbf{BT-O} & I & \textbf{30.1 {\scriptsize ± 1.50}} & \textbf{25.8 {\scriptsize ± 1.43}} & 19.4 {\scriptsize ± 1.29} & 48.6 {\scriptsize ± 1.63} & 49.8 {\scriptsize ± 1.55} &  41.1 {\scriptsize ± 1.61} & 48.9 {\scriptsize ± 1.63} & 44.60 \\
\textbf{TB} & I & 15.1 {\scriptsize ± 1.17} & 14.8 {\scriptsize ± 1.16} & \textbf{36.1 {\scriptsize ± 1.57}} & 46.1 {\scriptsize ± 1.63} & 48.8 {\scriptsize ± 1.63} &  \textbf{52.4 {\scriptsize ± 1.63}} & 52.2 {\scriptsize ± 1.63} & 46.73 \\
\midrule
\textbf{TB} & I+L & -& - & - & 46.2 {\scriptsize ± 1.63} & 49.0 {\scriptsize ± 1.63} &  \textbf{52.4 {\scriptsize ± 1.63}} & 52.3 {\scriptsize ± 1.63} & 49.77 \\
\textbf{BT-B} & I+L &- & - & - & \textbf{61.9 {\scriptsize ± 1.58}} & \textbf{64.2 {\scriptsize ± 1.56}} & 45.2 {\scriptsize ± 1.63} & \textbf{63.2 {\scriptsize ± 1.57}} & \textbf{58.14} \\
\textbf{BT-O} & I+L &- & - & - & 48.8 {\scriptsize ± 1.63} & 51.7 {\scriptsize ± 1.63} &  40.8 {\scriptsize ± 1.60} & 50.5 {\scriptsize ± 1.63} & 47.98 \\
\textbf{BT-B} & I+E &- & - & - & 25.9 {\scriptsize ± 1.43} & 26.1 {\scriptsize ± 1.43} &  33.1 {\scriptsize ± 1.61} & 30.1 {\scriptsize ± 1.50} & 28.65 \\
\textbf{BT-O} & I+E & -& - & - & 20.9 {\scriptsize ± 1.33} &  22.3 {\scriptsize ± 1.36} &  30.1 {\scriptsize ± 1.50} & 24.5 {\scriptsize ± 1.41} & 24.48 \\
\bottomrule
\end{tabularx}
}
\end{table}

\section{Conclusion}
\label{sections:conclusion}
We introduce \textsc{CrypticBio}, the largest publicly available multimodal dataset designed to better understand cryptic biodiversity and ultimately, accelerate trustworthy AI solutions for biodiversity. 
Curated from research-grade sources, this dataset focuses on visually confusing species groups with rich associated metadata, surpassing existing cryptic species datasets in scale by several orders of magnitude.
Our benchmarking across \textsc{CrypticBio}  underscore the value of context-aware multimodal datasets for advancing foundation models in biodiversity research, particularly for challenging cryptic species. 
\textsc{CrypticBio} is openly available (for download and browsing) on \href{https://huggingface.co/datasets/gmanolache/CrypticBio}{\textcolor{blue}{HuggingFace Datasets}}, and we release a comprehensive pipeline (\textsc{CrypticBio-Curate} on \href{https://github.com/georgianagmanolache/crypticbio/tree/master/crypticbio_curate}{\textcolor{blue}{GitHub}}) to facilitate custom subset creation and reproducibility.
With \textsc{CrypticBio}, we aim to accelerate the development of biodiversity AI models that are equipped to handle the real-world nuanced and context-dependent challenges of species ambiguity.


\bibliographystyle{unsrt}


\newpage
\appendix
\section{Ethics statement}
\label{supplementary_material:ethics_statement}
\subsection{Taxon selection}
\begin{figure}[H]
    \centering
   \includegraphics[width=\linewidth]{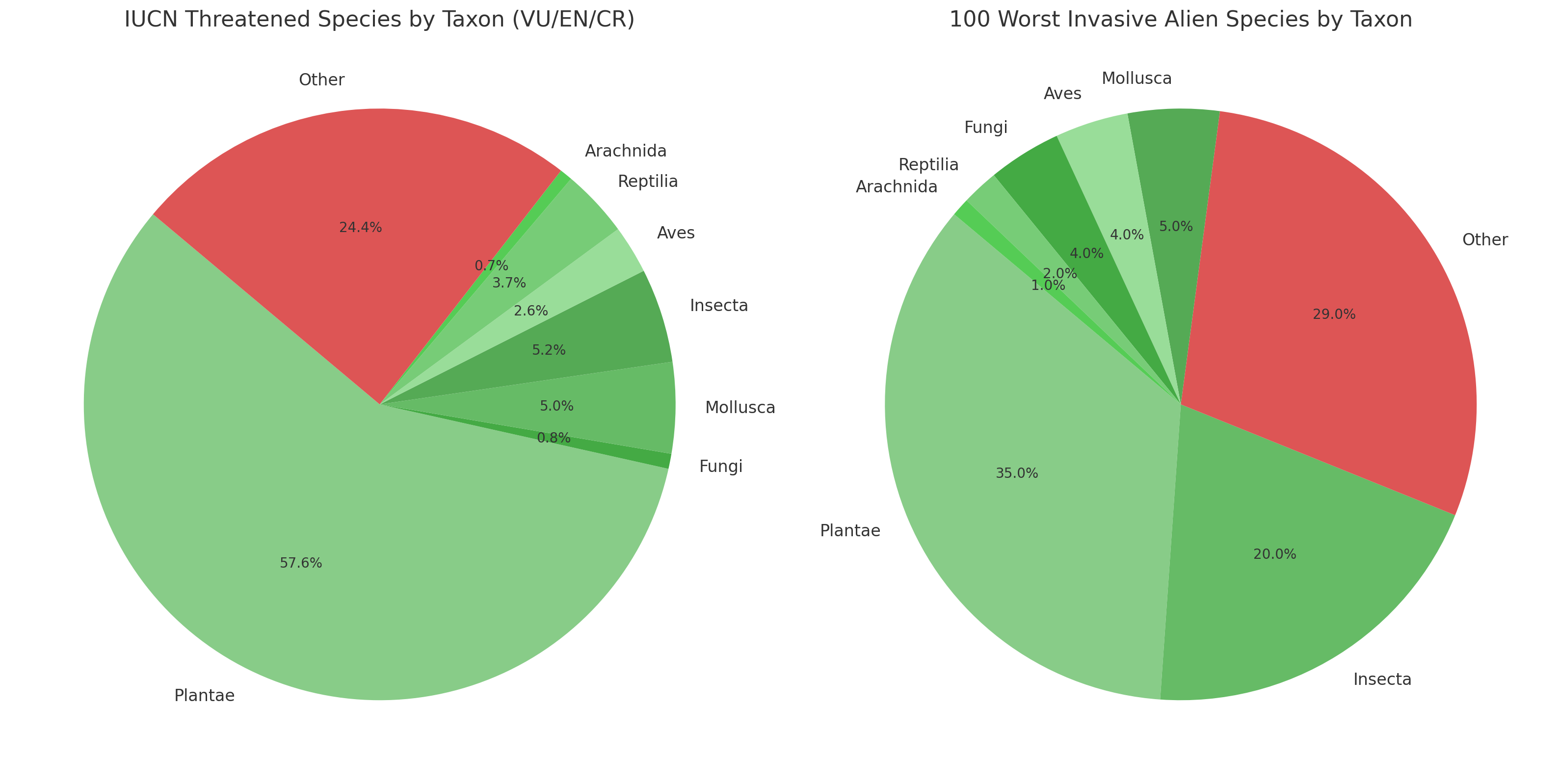}
    \caption{Taxa representativeness in   biodiversity conservation and policy change supervision: (left) IUCN~\cite{iucn2024} endangered species distribution (labeled VU=vulnerable, EN=endangered, CR=critically endangered, the highest threats); (right) GISD~\cite{IUCNGISD2013invasive} 100 worst alien species distribution.}
    \label{fig:taxon_distribution}
\end{figure}

\begin{figure}[t]
    \centering
\includegraphics[width=\textwidth]{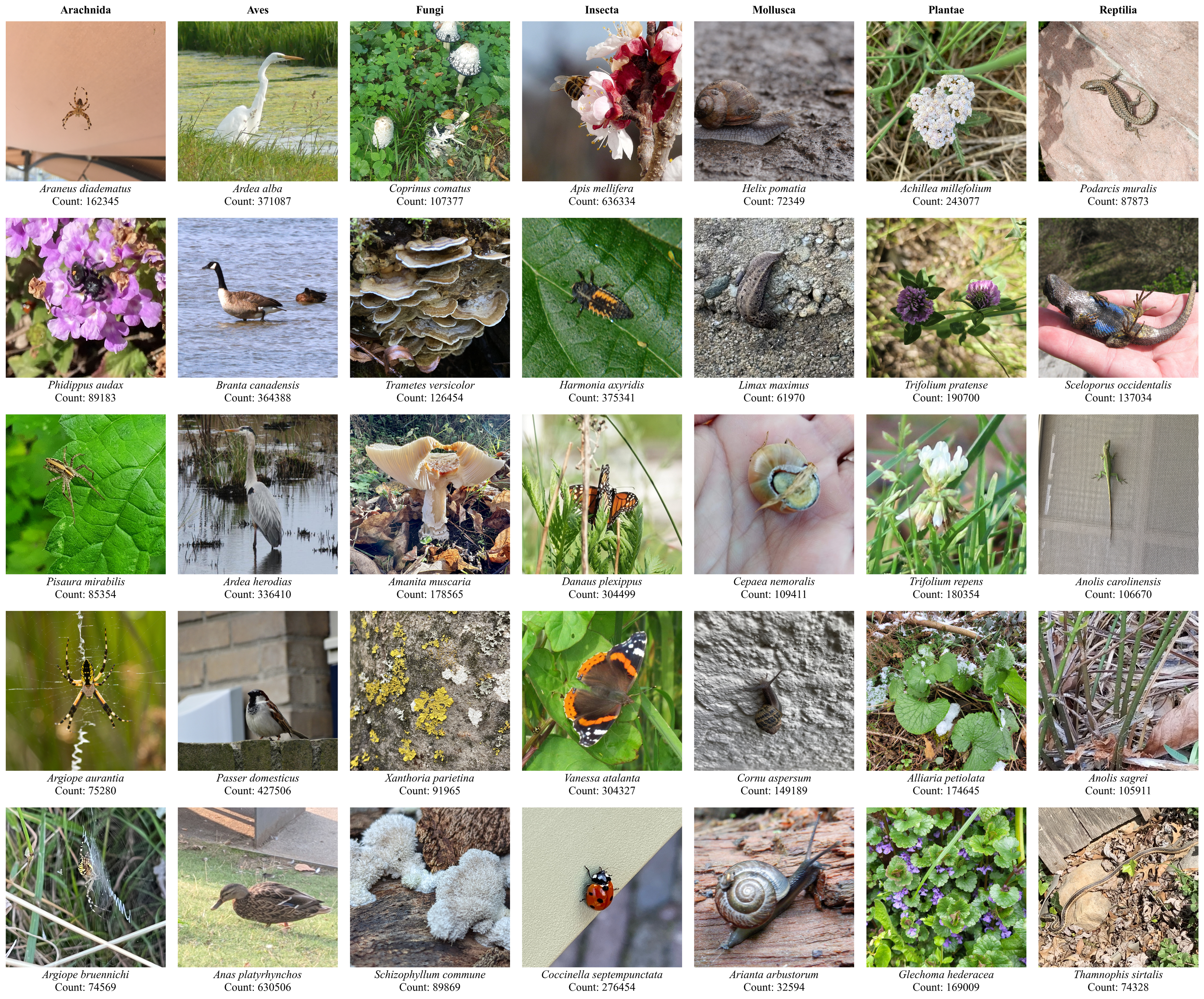}
    \caption{Examples of top five most frequent species and their counts in \textsc{CrypticBio}.}
    \label{fig:top_5_species}
\end{figure}

We select seven most representative taxa in biodiversity conservation and policy change supervision: \textit{Arachnida}, \textit{Aves}, \textit{Fungi}, \textit{Insecta}, \textit{Mollusca}, \textit{Plantae}, \textit{Reptilia}, as shown in Figure \ref{fig:taxon_distribution}.
These taxa represent the majority (>70\%) of threatened species (left) and harmful invaders (right), underscoring their significant ecological and economic impact. Figure \ref{fig:top_5_species} shows examples of top five most frequent species and their counts.

\subsection{Endangered species location disclosure}

It is critically important not to disclose the precise locations of threatened species because doing so can inadvertently put them at even greater risk. 
Many vulnerable species face threats from poaching, illegal wildlife trade, habitat disturbance, and over-collection. 
Sharing exact geographic coordinates, especially online or in open databases, can make it possible to locate and exploit these species.

\begin{figure}[H]
    \centering
    \begin{subfigure}[b]{0.48\textwidth}
        \centering
        \includegraphics[width=\linewidth]{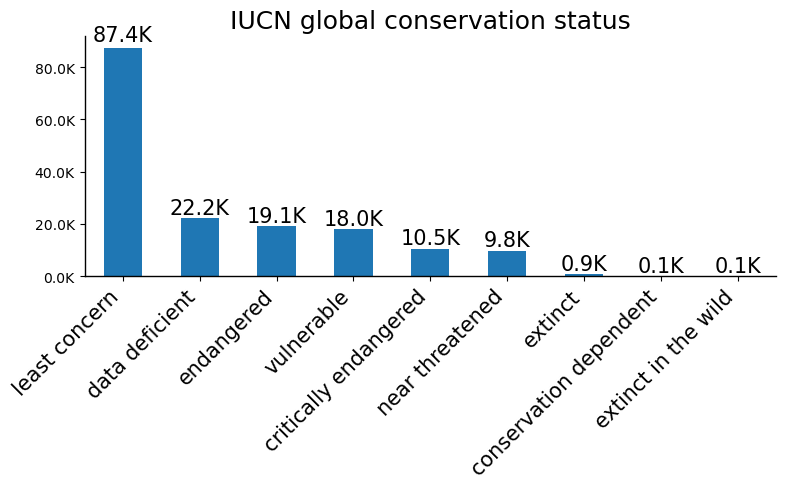}
    \end{subfigure}
    \hfill
    \begin{subfigure}[b]{0.48\textwidth}
        \centering
        \includegraphics[width=\linewidth]{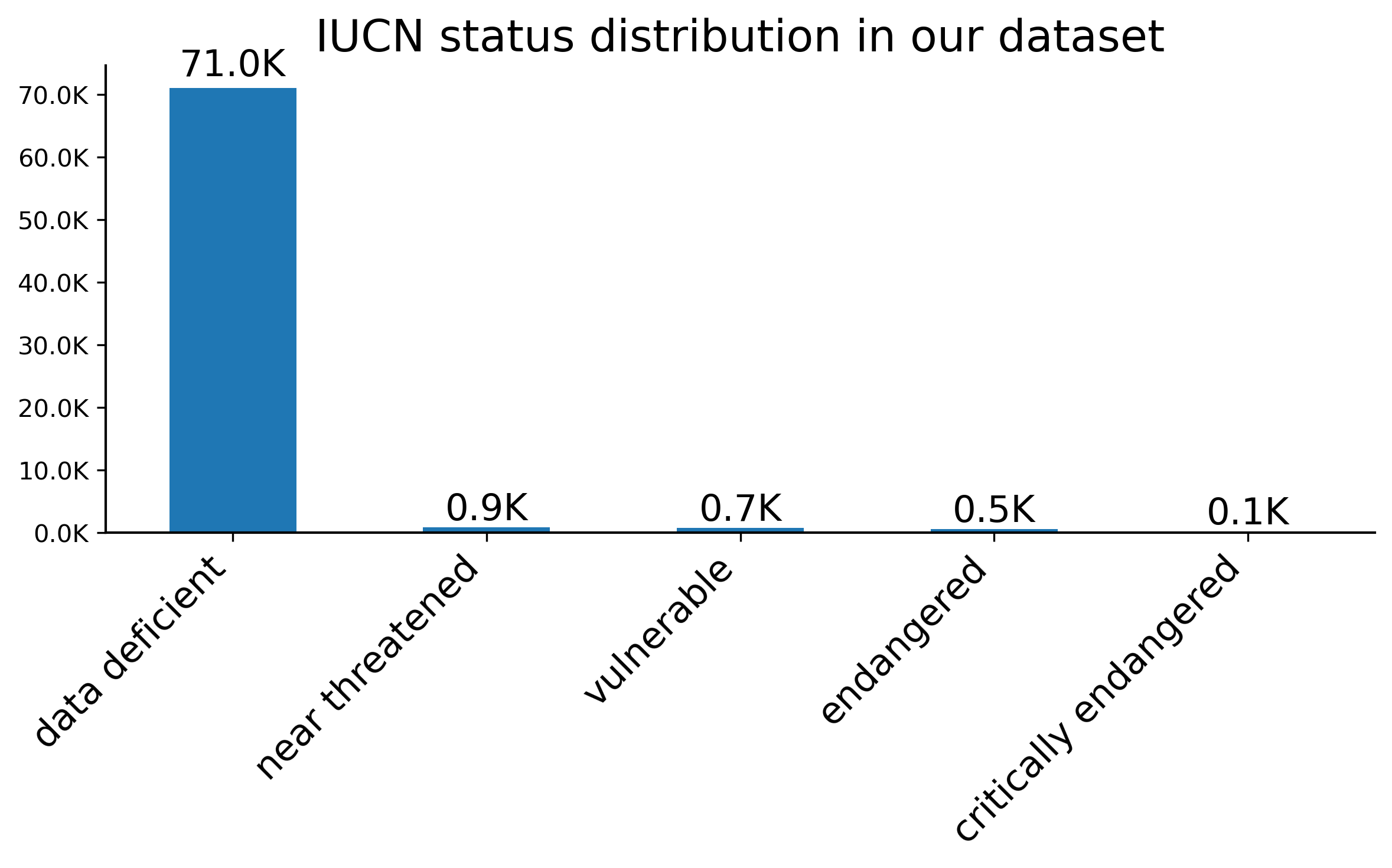}
    \end{subfigure}
    \caption{IUCN endangered species distribution: (left) IUCN endangered species distribution, (right) IUCN endangered species distribution in our dataset.}
    \label{fig:iucn_endangered_species_distribution}
\end{figure}

To mitigate the risks associated with the disclosure of sensitive biodiversity data, citizen science platforms iNaturalist and Observation.org implement automatic geoprivacy measures for taxa listed on the global IUCN (International Union for Conservation of Nature) Red List of Threatened Species~\cite{iucn2024}. 
Our dataset contains less than 2.3K endangered species according to IUCN, as shown in Figure~\ref{fig:iucn_endangered_species_distribution}.  

\subsection{Invasive alien species require geographical context}

Accurate identification of visually similar species is critical, particularly when distinguishing between invasive and non-invasive taxa. Many invasive species closely resemble native or benign taxa. 
Additionally, geographic information plays a critical role in this process, as the impact of a species can vary by region—what is considered invasive in one area may be benign or even native in another.
Overall, distinguishing invasive species within the appropriate geographic context is a foundational step in safeguarding biodiversity and maintaining ecological resilience. 

\subsection{Bias in vernacular species terminology across diverse taxa}

While we acknowledge the importance of incorporating common species terms alongside scientific (i.e., Latin binomial) nomenclature to enhance model performance~\cite{stevens2024bioclip}, the exclusive reliance on English vernacular names risks marginalizing indigenous and non-Western terminologies. 
Moreover, English speaking cultures may have their regional bias as well. 
For instance, species \textit{Perisoreus canadensis} is commonly referred to as the \textit{Canada Jay} in Canada, while in the United States is referred to as \textit{Gray Jay}~\cite{luccioni2023bugs}.
Currently, datasets like \textsc{TreeOfLife-10M}~\cite{stevens2024bioclip} and \textsc{BioTrove}~\cite{yang2024biotrove} include only one version of a species's vernacular name. 
We believe integrating  \textbf{multicultural and multilingual common terminology} preserves ecological knowledge and equity, and increases inclusivity and cultural reach. Thus, we
emphasize on enriching species scientific terminology with all common terms from iNaturalist Taxonomy~\cite{inaturalist2025taxonomy} (see dataset details is shown in Section~\ref{supplementary_material:inaturalist_taxonomy}.) 

\begin{figure}[H]
    \centering        \includegraphics[width=\linewidth]{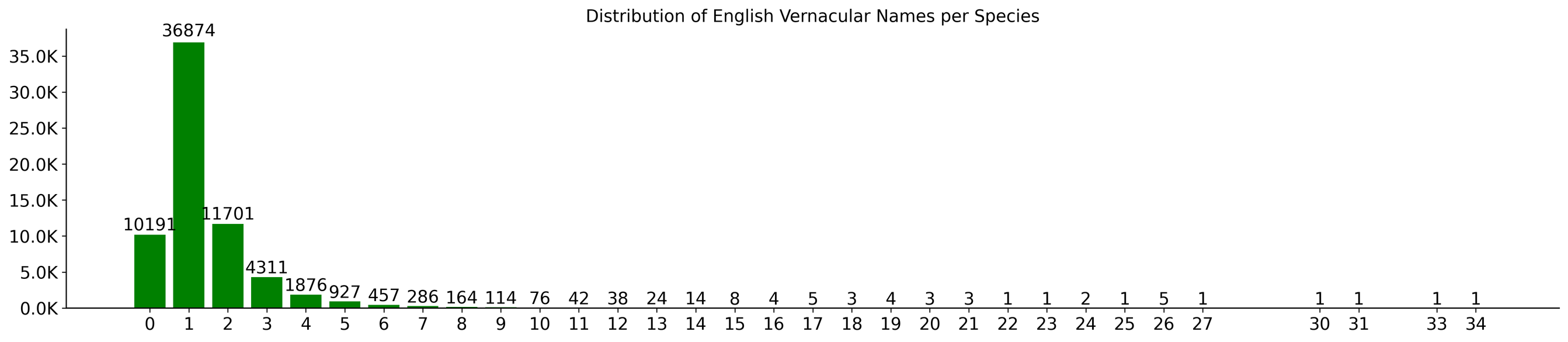}
    \caption{English vernacular names distribution in \textsc{CrypticBio} based on Naturalist Taxonomy~\cite{inaturalist2025taxonomy}.}
\label{fig:vernacular_name_distribution}
\end{figure}

Table~\ref{table:common_names_example} shows our dataset and comparable datasets \textsc{TreeOfLife-40M} and \textsc{CrypticBio} recorded English vernacular terminology for the widespread flower species \textit{Bellis perennis}. 
We include English vernacular names in \textsc{CrypticBio}, and provide a pipeline in \textsc{CrypticBio-Curate}
to enrich the dataset with language specific terminology. 
As illustrated in Figure~\ref{fig:vernacular_name_distribution}, approximately 30\% of species are associated with two or more English vernacular terms, whereas 15\% lack any recorded English terminology. It is worth noting that there are also species that have no vernacular names in English, which underlines the importance of preserving indigenous terminology.

\begin{table}[H]
\begin{center}
\caption{\textit{Bellis perennis} English vernacular names in existing biodiversity datasets and ours.}
\label{table:common_names_example}
\small
\begin{tabular}{l l} 
\toprule
\textbf{Dataset} & \textbf{Common name}\\
\midrule
\textsc{TreeOfLife-40M} & \textit{English daisy}\\
\textsc{BioTrove} & \textit{Lawn daisy}\\
\textsc{CrypticBio} & 
\textit{Common daisy}, \textit{English daisy}, \textit{Lawn daisy}\\
\bottomrule
\end{tabular}
\end{center}
\end{table}

\section{Dataset suite}
\label{supplementary_material:datasets_suite}
Table~\ref{table:datasets_suite} summaries all datasets used for the contruction of \textsc{CrypticBio}, while section \ref{supplementary_material:gbif_details}-\ref{supplementary_material:inaturalist_similar_species} detail each data source.

\begin{table}[H]
\begin{center}
\caption{Dataset suite used in the curation of \textsc{CrypticBio}.}
\label{table:datasets_suite}
\small
\begin{tabularx}{\textwidth}{p{4.5cm} X} 
\toprule
\textbf{Dataset} & \textbf{Description}\\
\midrule
GBIF~\cite{gbif2025} & Occurrence records including species observations with associated metadata such as date, location, and scientific name. Served as the primary source of biodiversity data. \\
GBIF Backbone Taxonomy~\cite{gbif2023backbone} & Taxonomic reference for resolving scientific names and aligning species-level classifications across datasets. \\
iNaturalist Taxonomy~\cite{inaturalist2025taxonomy} & Cross-referencing vernacular and scientific names, and refining taxonomic granularity particularly for user-contributed observations. \\
iNaturalist "Similar Species" &  Cryptic group composed of other species commonly misidentified with a focal species\\
\bottomrule
\end{tabularx}
\end{center}
\end{table}

\subsection{GBIF}
\label{supplementary_material:gbif_details}
GBIF~\cite{gbif2025} primarily aggregates research-grade biodiversity data, focusing on species occurrence records derived from scientific sources such as museum collections, academic research, and validated citizen science observations. As a result, the dataset emphasizes verifiable, expert-curated information rather than general public or commercial data.

GBIF uses the Darwin Core standard—a widely adopted vocabulary for sharing biodiversity data. Each GBIF dataset is typically a Darwin Core Archive, structured follows:

\begin{itemize}
    \item A core CSV file \texttt{occurrence.txt} with observation or specimen records. Each row in \texttt{occurrence.txt} is one occurrence (a species observation or specimen record).
    
    \item Optional extension files: \texttt{multimedia.txt}, \texttt{identification.txt}.

    \item A \texttt{meta.xml} file describing the structure.

    \item A \texttt{eml.xml} file with metadata about the dataset.
\end{itemize}

Table~\ref{table:gbif_details} summarizes fields used in creation of \textsc{ConfoundingBio}.

\textbf{Occurrences} GBIF's \texttt{occurrence.txt} file enumerates 223 fields, however, many fields are often empty. Original identifiers and provenance data files, such as dataset’s ID and name (iNaturalist Research-Grade Observations and Observation.org) and original record’s unique ID. One of the most important metadata fields is \texttt{basisOfRecord}, which tells what kind of occurrence the record is—for example, whether it is a direct human observation, a museum specimen, or machine-generated.

Extensive biological and taxonomic information enumerates full scientific name (usually with authorship), taxonomic level of the record (\texttt{taxonRank}), taxonomic hierarchy broken into separate fields (i.e., \texttt{kingdom}, \texttt{phylum}, \texttt{class}, \texttt{order}, \texttt{superfamily}, \texttt{family}, \texttt{tribe}, \texttt{subtribe}, \texttt{subfamily}, \texttt{genus}, \texttt{subgenus}) and common name (if provided).
An important field is \texttt{taxonomicStatus} which records the status of the observation (either species, genus, family) is  the currently valid/recognized name in taxonomy (marked as \texttt{ACCEPTED}) or a synonym and its usage is questionable, incorrectly used. We use only \texttt{ACCEPTED} \texttt{taxonomicStatus} of \texttt{taxonRank} species.

\begin{table}[t]
\begin{center}
\caption{GBIF core CSV files \texttt{occurrence.txt} and \texttt{multimedia.txt} essential field description.}
\label{table:gbif_details}
\small
\begin{tabularx}{\textwidth}{l X} 
\toprule
\textbf{Field} & \textbf{Description}\\
\midrule
\texttt{gbifID} & Unique identifier for occurrence records\\
\texttt{scientificName} & Species observation scientific name\\
\texttt{taxonRank} & Observation taxonomic level (i.e. species, genus, family) \\
\texttt{decimalLatitude}, 
 \texttt{decimalLongitude} & Geographic coordinates in decimals\\
\texttt{year}, \texttt{month}, \texttt{day} & Date parts (often included separately too)\\
\texttt{type} & The type of media available usually \texttt{StillImage}, \texttt{Sound}, or \texttt{MovingImage}.\\
\texttt{identifier} & Direct URL to raw media content (image/audio/video)\\
\texttt{license} & Data license (usually CC-BY or CC0)\\
\bottomrule
\end{tabularx}
\end{center}
\end{table}

The extensive geographic location fields describe where an organism was observed or collected. Core fields detail latitude and longitude coordinates in decimal degrees and radius of uncertainty around the point in meters (e.g. 30, meaning $\pm$30 meters). More locality details include country details and well as free-text description of the place also written in other languages than English.

Apart from \texttt{license}, the main field used for legal reuse, there may be detailed access and rights data, \texttt{accessRights}, \texttt{rightsHolder} giving more contextual info about the data accessibility than the strict license field, however, rarely populated in GBIF records. 

Other information less relevant for the modern biodiversity is geological context set of fields (13 fields) in GBIF’s, designed to describe the stratigraphic and temporal layers from which a fossil or subfossil specimen was recovered. These fields are especially important for paleontology, stratigraphy, and earth history research and is relevent when an observation is a fossil (i.e., \texttt{basisOfRecord} is  \texttt{FOSSIL\_SPECIMEN}).

Data quality fields are critical for assessing whether a record is usable, reliable, or problematic. We use \texttt{hasGeospatialIssues} boolean to filter all includes valid geographic coordinates. Another interesting field \texttt{iucnRedListCategory} categories taxon conservation status accoridng to International Union for Conservation of Nature (IUCN) Red List~\cite{iucn2024}, although this data is not consistent throughout the records.

\textbf{Multimedia} GBIF's \texttt{multimedia.txt} file enumerates 15 fields, and can be joined to \texttt{occurrence.txt} via \texttt{gbifID}. These fields provide access to the media itself (i.e., \texttt{identifier}) and its context (i.e., \texttt{type}), and specify who created or owns the media, and how it can be used (i.e., \texttt{license}). 

\subsection{GBIF Backbone Taxonomy}
\label{supplementary_material:gbif_taxonomy}
GBIF Backbone Taxonomy~\cite{gbif2023backbone} is structured as a Darwin Core Archive, as follows:

\begin{itemize}

    \item A core TSV (Tab-Separated Values)  file \texttt{Taxon.tsv}  the primary taxonomic information.
    
    \item  Extension files: 
    \begin{itemize}
        \item \texttt{VernacularName.tsv} provides common names (vernacular names) for taxa.
        \item \texttt{TypeAndSpeciment.tsv} lists information about taxonomic identifications of species. 
        \item \texttt{Description.tsv} contains textual descriptions of taxa, offering additional information such as morphology, behavior, or ecology.
        \item \texttt{Distribution.tsv} provides geographic and ecological information associated with specific taxa.
        \item \texttt{Reference.tsv} lists bibliographic references related to the taxa.
        \item \texttt{Multimedia.tsv} links media resources, such as images or sounds, to taxa.
    \end{itemize}

    \item A \texttt{meta.xml} file describing the structure.

    \item A \texttt{eml.xml} file with metadata about the dataset.
\end{itemize}

\begin{table}[H]
\begin{center}
\caption{GBIF Backbone Taxonomy \texttt{Taxon.tsv} essential field description.}
\label{table:gbif_taxonomy_details}
\small
\begin{tabularx}{\textwidth}{l X} 
\toprule
\textbf{Field} & \textbf{Description}\\
\midrule
\texttt{canonicalName} & Unique species scientific name (Latin binomial), lowest taxonomic rank \\
\texttt{kingdom} & Highest taxonomic rank (Latin uninomial)\\
\texttt{phylum} &  Second taxonomic rank (Latin uninomial)\\
\texttt{class} & Third taxonomic rank (Latin uninomial)\\
\texttt{order} &  Forth taxonomic rank (Latin uninomial)\\
\texttt{family} & Fifth taxonomic rank (Latin uninomial)\\
\texttt{genus} & Sixth taxonomic rank (Latin uninomial)\\
\bottomrule
\end{tabularx}
\end{center}
\end{table}

\begin{table}[b]
\caption{Diversity in different taxonomy levels in GBIF Backbone Taxonomy~\cite{gbif2023backbone} (left) and \textsc{CrypticBio} (right).}
\label{table:gbif_backbone_taxonomy}
\noindent
\small
\begin{minipage}{0.5\textwidth}
  \centering
  \begin{tabular}{c c}
    \toprule
    \textbf{Level} & \textbf{Count} \\
    \midrule
    kingdom & 8 \\
    phylum & 169 \\
    class & 519 \\
    order & 1953 \\
    family & 15139 \\
    genus & 268644 \\
    species & 3389404\\
    \bottomrule
  \end{tabular}
\end{minipage}
\hfill
\begin{minipage}{0.5\textwidth}
  \centering
  \begin{tabular}{c c}
    \toprule
    \textbf{Level} & \textbf{Count} \\
    \midrule
 kingdom & 3 \\
phylum & 14 \\
class & 56 \\
order & 351 \\
family & 2036 \\
genus & 17327 \\
species & 67140 \\
    \bottomrule
  \end{tabular}
\end{minipage}
\end{table}

GBIF observations enumerate taxonomic hierarchy of 11 levels (\texttt{kingdom}, \texttt{phylum}, \texttt{class}, \texttt{order}, \texttt{superfamily}, \texttt{family}, \texttt{tribe}, \texttt{subtribe}, \texttt{subfamily}, \texttt{genus}, \texttt{subgenus}) broken into separate fields, however, many fields are often empty. Instead, we use GBIF's Backbone Taxonomy~\cite{gbif2023backbone} to enrich observations at \texttt{species} taxonomic level with six taxonomic hierarchy levels: \texttt{kingdom}, \texttt{phylum}, \texttt{class}, \texttt{order}, \texttt{family}, \texttt{genus}. Diversity in different taxonomy levels in  GBIF Backbone Taxonomy~\cite{gbif2023backbone} and \textsc{CrypticBio} is shown in Table~\ref{table:gbif_backbone_taxonomy}.
Unlike the selected six taxonomic hierarchy levels, levels like \texttt{superfamily}, \texttt{tribe}, \texttt{subtribe},  \texttt{subfamily},\texttt{subgenus} not consistently recorded in taxonomy~\cite{iczn1999}.

\subsection{iNaturalist Taxonomy}
\label{supplementary_material:inaturalist_taxonomy}
iNaturalist Taxonomy~\cite{inaturalist2025taxonomy} is structured as a Darwin Core Archive, as follows:

\begin{itemize}

    \item A core CSV file \texttt{taxa.csv}  the primary taxonomic information.
    
    \item  Extension files: vernacular names CSV files for each language, encoded as \texttt{VernacularNames-[language].csv}; there are 1091 language specific CSV files.

    \item A \texttt{meta.xml} file describing the structure.

    \item A \texttt{eml.xml} file with metadata about the dataset.
\end{itemize}

\begin{table}[t]
\begin{center}
\caption{iNaturalist Taxonomy \texttt{taxa.csv} and \texttt{VernacularNames-english.csv} essential field description.}
\label{table:inaturalist_taxonomy_details}
\small
\begin{tabularx}{\textwidth}{l X} 
\toprule
\textbf{Field} & \textbf{Description}\\
\midrule
\texttt{id} & Unique identifier for occurrence records\\
\texttt{scientificName} & Species observation scientific name (i.e. Latin binomial)\\
\texttt{vernacularName} & Common or vernacular name (e.g., "Lawn daisy", "English daisy")\\
\texttt{language} & Language of the vernacular name encoded with ISO 639 standard (e.g., en, es, fr)\\
\bottomrule
\end{tabularx}
\end{center}
\end{table}

We include all vernacular terminology in \texttt{VernacularNames-english.csv}. We provide a pipeline in \textsc{CrypticBio-Curate} to enrich the dataset with language specific terminology.

\subsection{iNaturalist Similar Species}
\label{supplementary_material:inaturalist_similar_species}
The "Similar Species" feature on iNaturalist is designed to assist annotators in distinguishing between species that are often confused due to their similar appearances. This tool is particularly useful for species which share visual characteristics with other species.

\begin{figure}[H]
    \centering
        \includegraphics[width=0.75\linewidth]{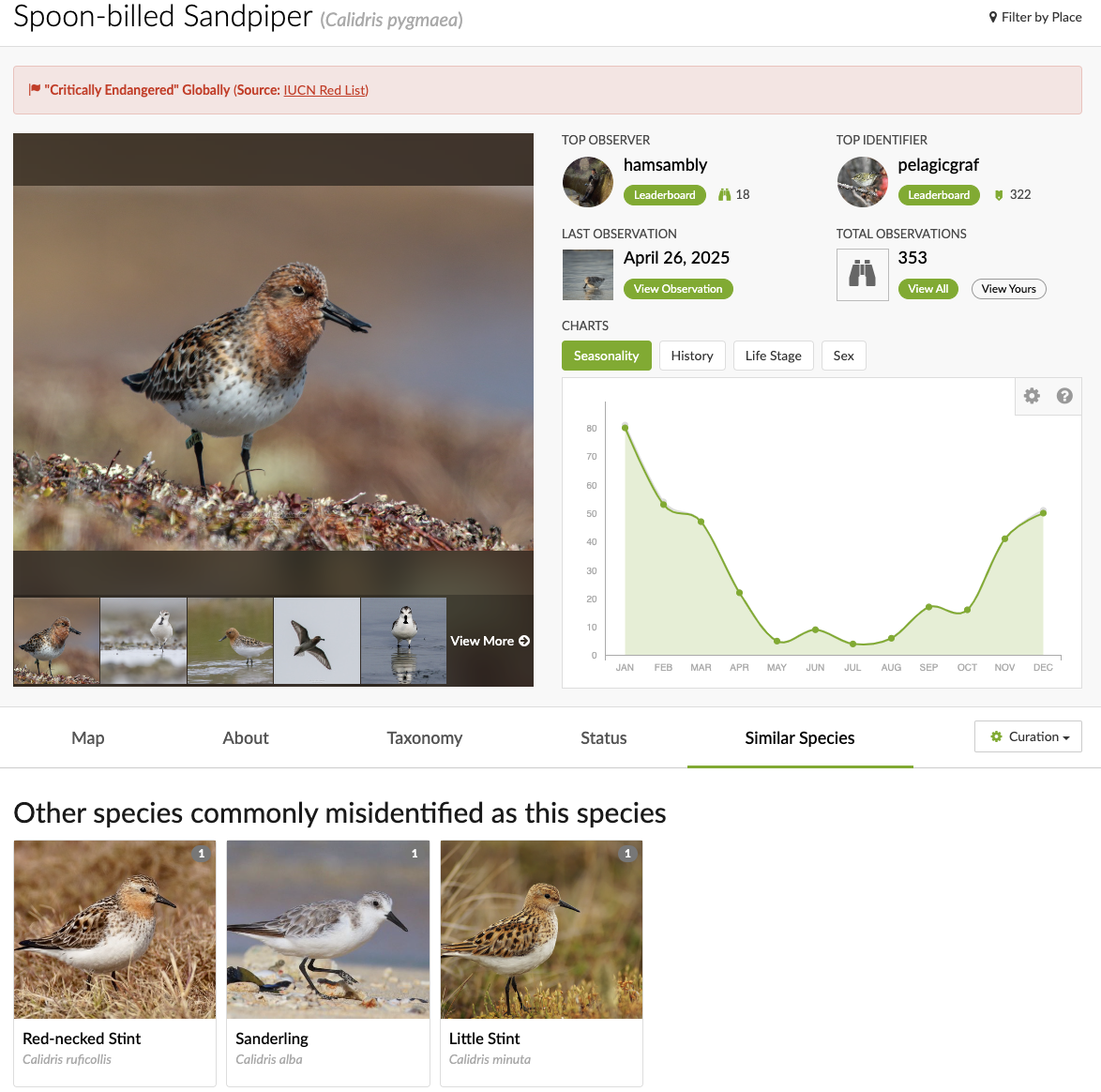}
  
    \caption{iNaturalist "Similar Species"  tab for \textit{Calidris pygmaea}.}
    \label{fig:inaturalist_similar_species}
\end{figure}

The feature relies on the collective input of the iNaturalist community. When users submit observations and identifications, the system tracks instances where species are misidentified, helping to build the "Similar Species" list. An example of such list is shown in Figure \ref{fig:inaturalist_similar_species} for species \textit{Calidris pygmaea}.

This feature is not always visible for all species. Its presence depends on the availability of sufficient observation data and a documented history of misidentifications between the focal species and others. For many less-observed, rare, or underrepresented taxa, the feature may not appear at all. This is because the system relies entirely on community-driven data and automated algorithms that detect patterns in user identifications; it is not manually curated. As a result, even if a species has close look-alikes, the "Similar Species" tab may be absent if those confusions have not been frequently recorded by users. This limitation is especially noticeable for obscure species or those from poorly documented regions, underscoring the importance of consulting external field guides or expert communities when the feature is not available.

\section{\textsc{CrypticBio} dataset}
\label{supplementary_material:dataset_details}
\begin{figure}[H]
    \centering
\includegraphics[width=\textwidth]{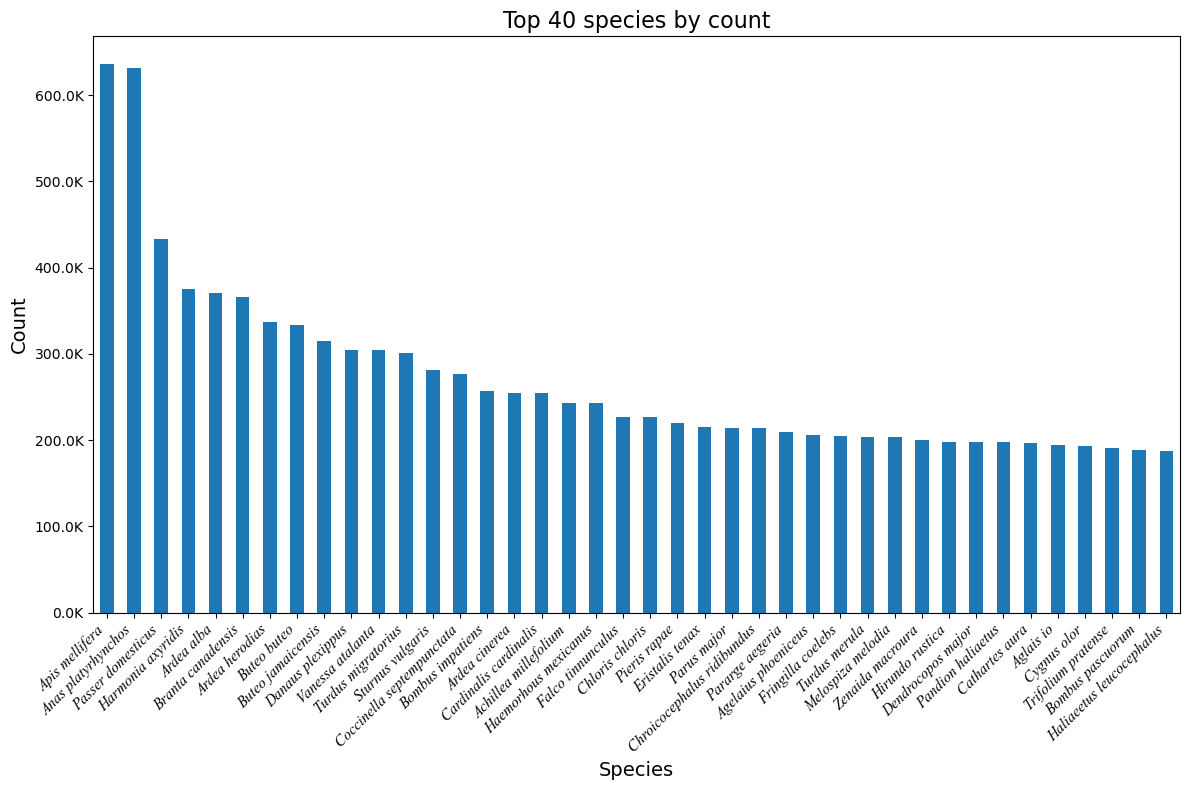}
    \caption{The 40 most frequent species.}
    \label{fig:top_40_species}
\end{figure}

\begin{figure}[H]
    \centering
\includegraphics[width=\textwidth]{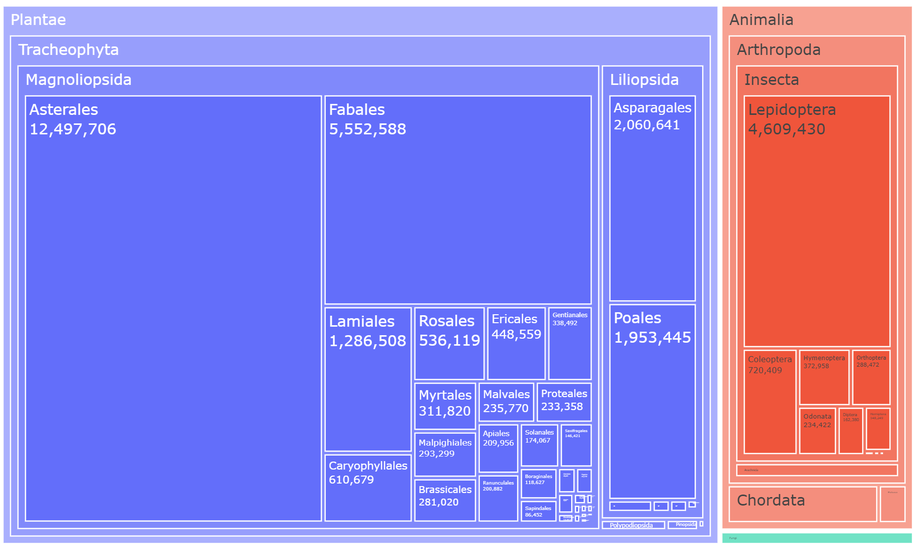}
    \caption{Treemap diagram, starting from \textit{kingdom}. The nested boxes represent \textit{phyla}, \textit{classes}, \textit{orders}, and \textit{families}. Box size represents the relative number of samples in the dataset.}
    \label{fig:tree}
\end{figure}

Figure~\ref{fig:top_40_species} displays CrypticBio top 40 most frequent species, Figure \ref{fig:tree} shows the treemap diagram, from \textit{kingdom}, \textit{phyla}, \textit{classes},
\textit{orders}, and \textit{families}, and Table~\ref{table:comparisons_all} shows comparable datasets.

\begin{table}[H]
\caption{\textsc{CrypticBio} comparable datasets and benchmarks.}
\label{table:comparisons_all}
\centering
{\scriptsize 
\begin{tabular}{ c  c  c  c  c  c } 
\toprule
\textbf{Datasets} &  \textbf{Images}& \textbf{Species} & \textbf{Annotations} & \textbf{Source} & \textbf{Features}\\
\midrule
\makecell{\textbf{\textsc{CrypticBio}}} & 166.0M & 71.0K & \makecell{common (multicultural and   \\ multilingual), scientific terms, \\
taxonomic hierarchies, location, \\ date, confounding species \\ groups} & \makecell{ GBIF (iNaturalist and \\Observation.org), \\GBIF Backbone \\ Taxonomy~\cite{gbif2023backbone}, \\ iNaturalist \\Taxonomy~\cite{inaturalist2025taxonomy} } & \makecell{multimodal, \\data-driven \\ cryptic \\ species groups \\ (52K groups)}\\
\midrule
\makecell{\textbf{\textsc{BioTrove}~\cite{yang2024biotrove}}}  & 161.9M  & 366.6K & \makecell{common, scientific terms,\\ taxonomic hierarchies}& iNaturalist & \makecell{biased common \\ species terminology  \\ annotations}\\
\midrule
\makecell{\textbf{\textsc{TreeOfLife-10M}~\cite{stevens2024bioclip}}}  & 10.4M  & 454.1K & \makecell{common, scientific terms,\\ taxonomic hierarchies} & \makecell{ iNaturalist,\\ Encyclopedia \\of Life (EOL)\cite{eol2025},\\ \textsc{BioScan-1M}\cite{gharaee2023step}} &  \makecell{biased common \\ species terminology \\ annotations}\\
\hline
\makecell{\textbf{\textsc{TaxaBind-8K}~\cite{sastry2024taxabind}}}  & 8.8K  & 2.2K & \makecell{common, scientific term, \\taxonomic hierarchies, location,\\ environmental features,\\ audio recordings, satellite \\ imagery} & \makecell{iNaturalist, \\ iNat2021\cite{van2021benchmarking}, \\ Santinel-2\cite{s2maps2023}, \\ WorldClim-2.1\cite{fick2017worldclim} }& multimodal\\
\midrule
\makecell{\textbf{\textsc{Bumble Bees}~\cite{spiesman2021assessing}}\\ (not publicly available)} & 89K & 36 & \makecell{scientific terms} & \makecell{iNaturalist,\\Bumble Bee \\Watch~\cite{hatfield2024bumblebeewatch}, \\ BugGuide~\cite{bugguide2025}} & \makecell{manually selected \\cryptic \\ species group \\ (1 group)}\\
\hline
\makecell{\textbf{\textsc{Turtles}~\cite{baek2024turtles}} \\(not publicly available)} & 6.9K & 36 & \makecell{common, scientific terms} & \makecell{Internet} & \makecell{manually selected \\cryptic \\ species group \\ (1 group)} \\
\midrule
\makecell{\textbf{\textsc{Amazon Parrots}~\cite{kim2025parrots}}} & 14K & 35 & \makecell{scientific terms} & \makecell{iNaturalist,\\ eBird~\cite{ebird2021}, \\Google Images} & \makecell{manually selected \\cryptic \\ species group \\ (16 groups)}\\
\hline
\makecell{\textbf{\textsc{Confounding}} \\ \textbf{\textsc{Species}~\cite{chiranjeevi2023deep}}\\ (not publicly available)}  & 100 & 10 & \makecell{scientific term, confounding \\ species pairs}& iNaturalist & \makecell{manually selected \\ cryptic \\ species pairs}\\
\midrule
\makecell{\textbf{\textsc{Squamata}}\\ \textbf{\textsc{Lacertidae}} \\ \textbf{\textsc{Podarcis}~\cite{pinho2022squamata}}}  & 4.0K & 9 & \makecell{scientific terms} & \makecell{personal collection \\ during field surveys} & \makecell{manually selected \\cryptic \\species  group  \\ (1 group)}\\
\midrule
\makecell{\textbf{\textsc{Chiroptera}}\\ \textbf{\textsc{Rhinolophidae}} \\ \textbf{\textsc{Rhinolophus}~\cite{cao2024bats}}} & 293 & 7 & \makecell{scientific terms} & \makecell{personal collection \\ during field surveys} & \makecell{manually selected \\cryptic \\ species group \\ (1 group)} \\
\bottomrule
\end{tabular}
}
\end{table}


\section{New benchmarks}
\label{supplementary_material:new_benchmarks}
\subsection{\textsc{CrypticBio-Common} benchmark details}
\begin{table}[H]
\begin{center}
\caption{\textsc{CrypticBio-Common} subset distribution}
\label{table:common_stats}
\small
\begin{tabular}{c c c c} 
\toprule
\textbf{Taxon} & \textbf{Selected species} & \textbf{\#Associated cryptic species} & \textbf{\#Observations}\\
\midrule
(\textit{Arachnida}) & \textit{Parasteatoda tepidariorum} & 24 & 500425\\
(\textit{Aves}) & \textit{Passer domesticus} & 25 & 2836119\\
(\textit{Fungi}) & \textit{Amanita muscaria} & 24 & 333551\\
(\textit{Insecta}) & \textit{Harmonia axyridis} & 25 & 947893\\ 
(\textit{Mollusca}) & \textit{Cornu aspersum} & 25 & 542833\\
(\textit{Plantae}) & \textit{Bellis perennis} & 24 & 839657\\
(\textit{Reptilia}) & \textit{Zootoca vivipara} & 19 & 385535 \\
\bottomrule
\end{tabular}
\end{center}
\end{table}

\begin{figure}[H]
    \centering
    \includegraphics[width=\textwidth]{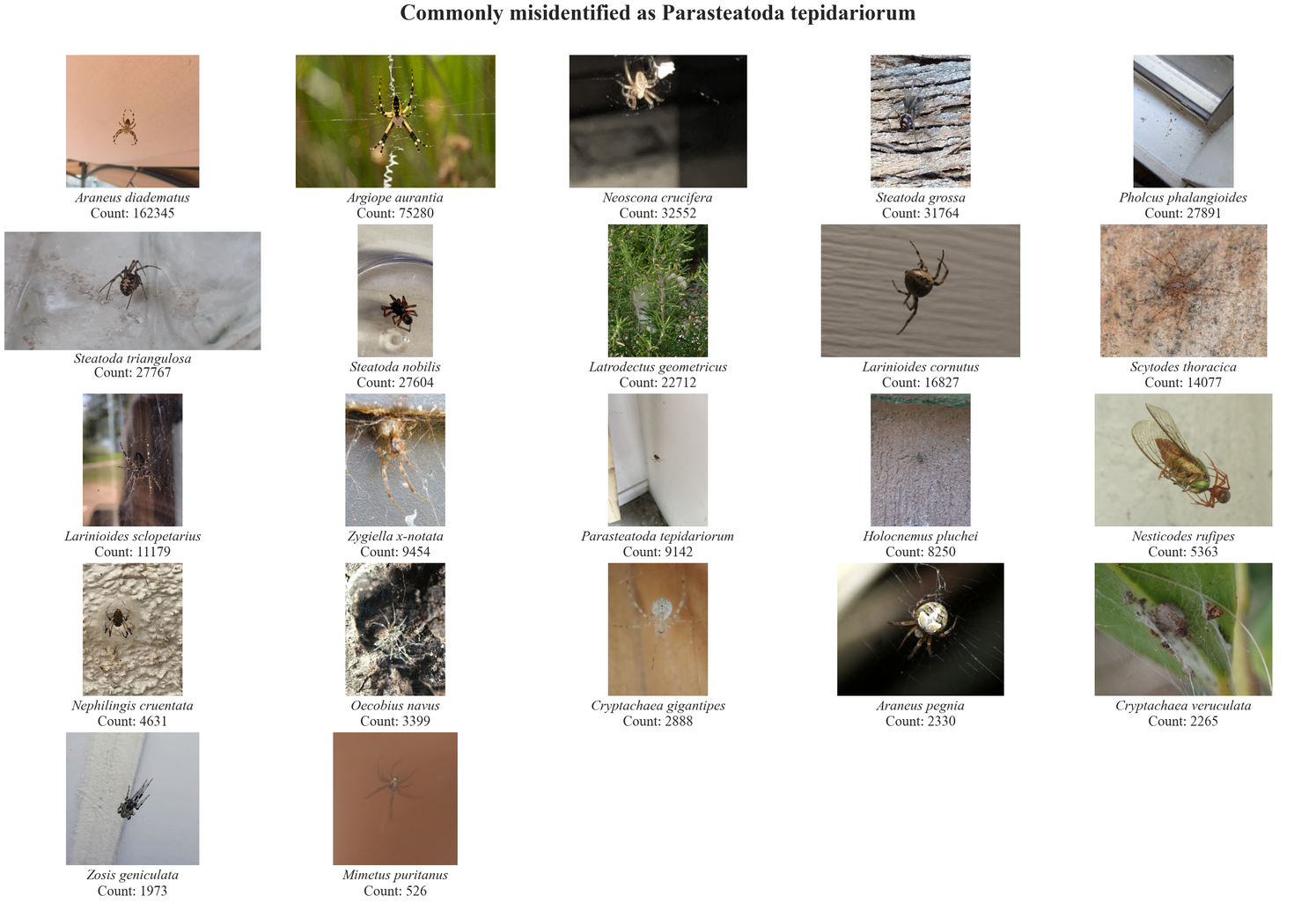}
    \caption{Sample of commonly misidentified of selected species (\textit{Arachnida}) \textit{Parasteatoda tepidariorum}.}
    \label{fig:Parasteatoda_tepidariorum}
\end{figure}

\begin{figure}[H]
    \centering
    \includegraphics[width=\textwidth]{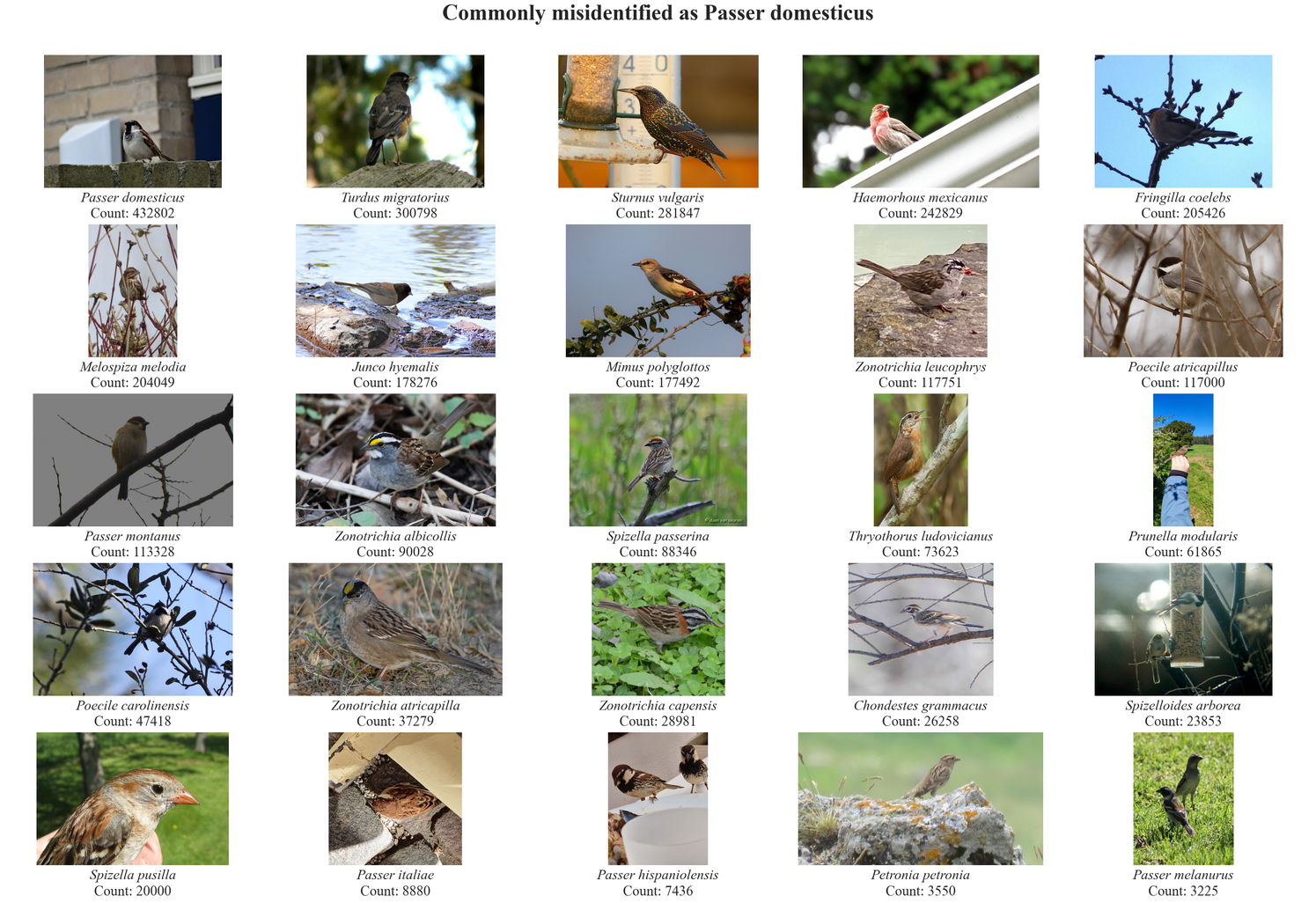}
    \caption{Sample of commonly misidentified of selected species (\textit{Aves}) \textit{Passer domesticus}.}
    \label{fig:Passer_domesticus}
\end{figure}

\begin{figure}[H]
    \centering
    \includegraphics[width=\textwidth]{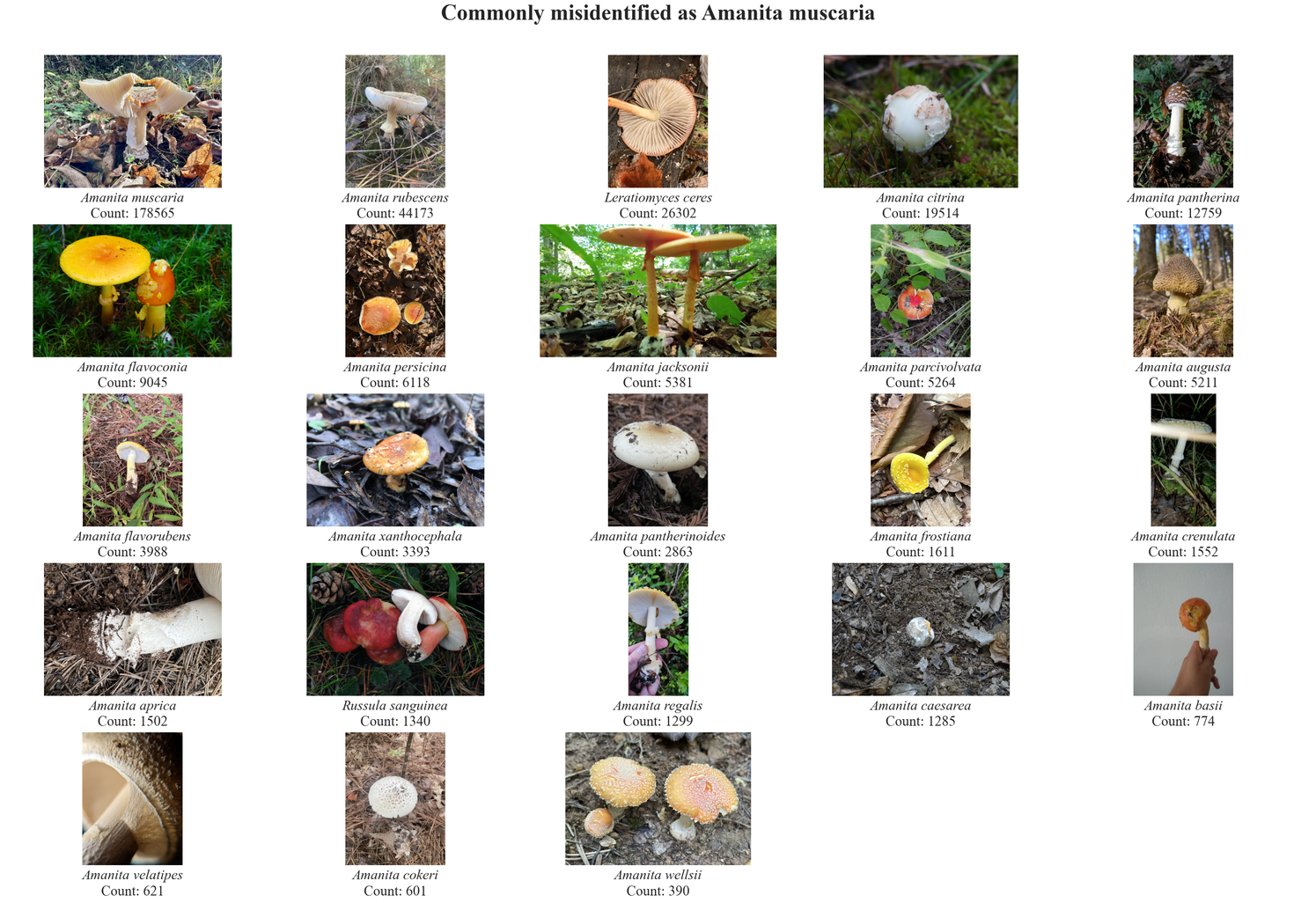}
    \caption{Sample of commonly misidentified of selected species (\textit{Fungi}) \textit{Amanita muscaria}.}
    \label{fig:Amanita_muscaria}
\end{figure}

\begin{figure}[H]
    \centering
    \includegraphics[width=\textwidth]{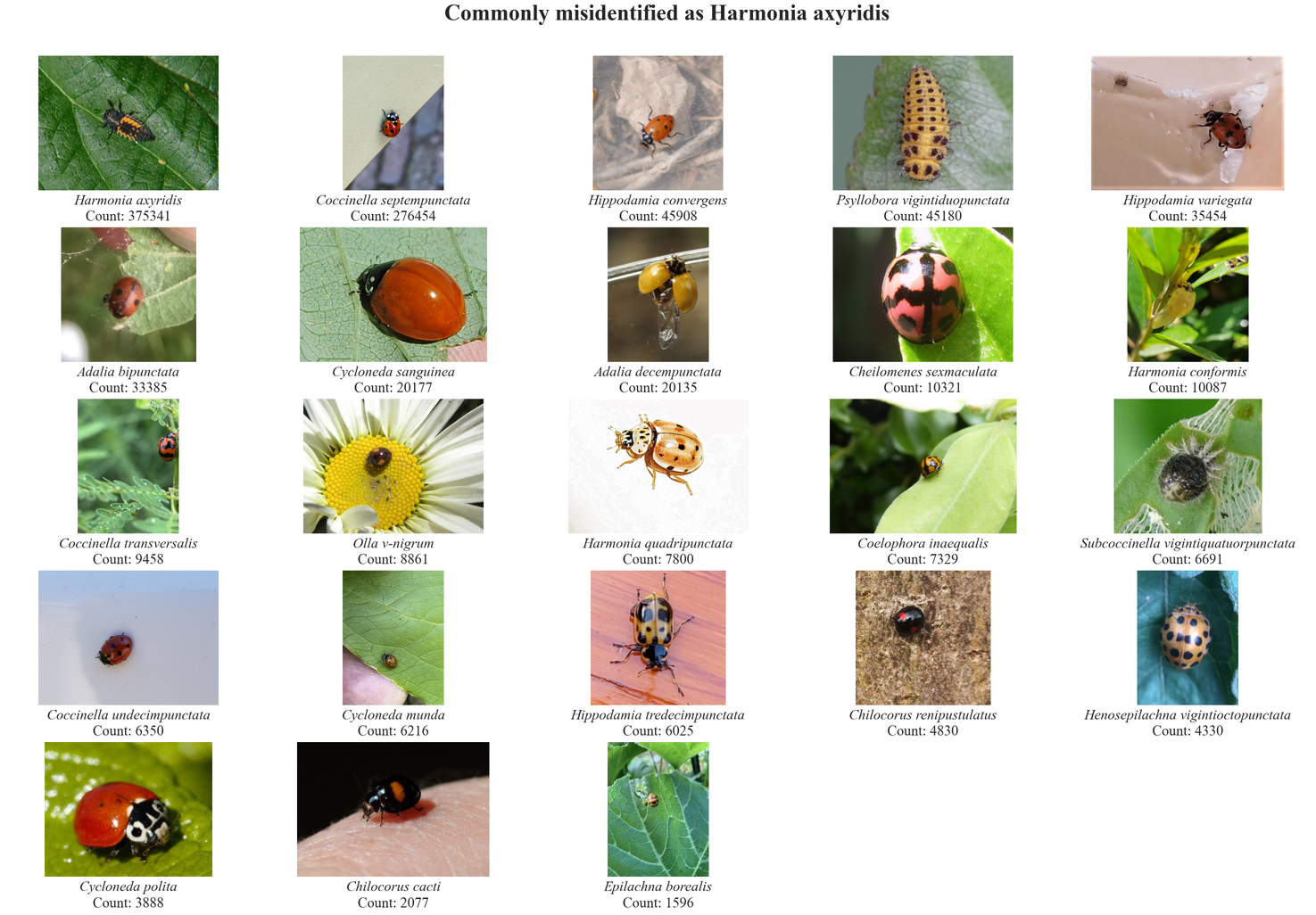}
    \caption{Sample of commonly misidentified of selected species (\textit{Insecta}) \textit{Harmonia axyridis}.}
    \label{fig:Harmonia_axyridis}
\end{figure}

\begin{figure}[H]
    \centering
    \includegraphics[width=\textwidth]{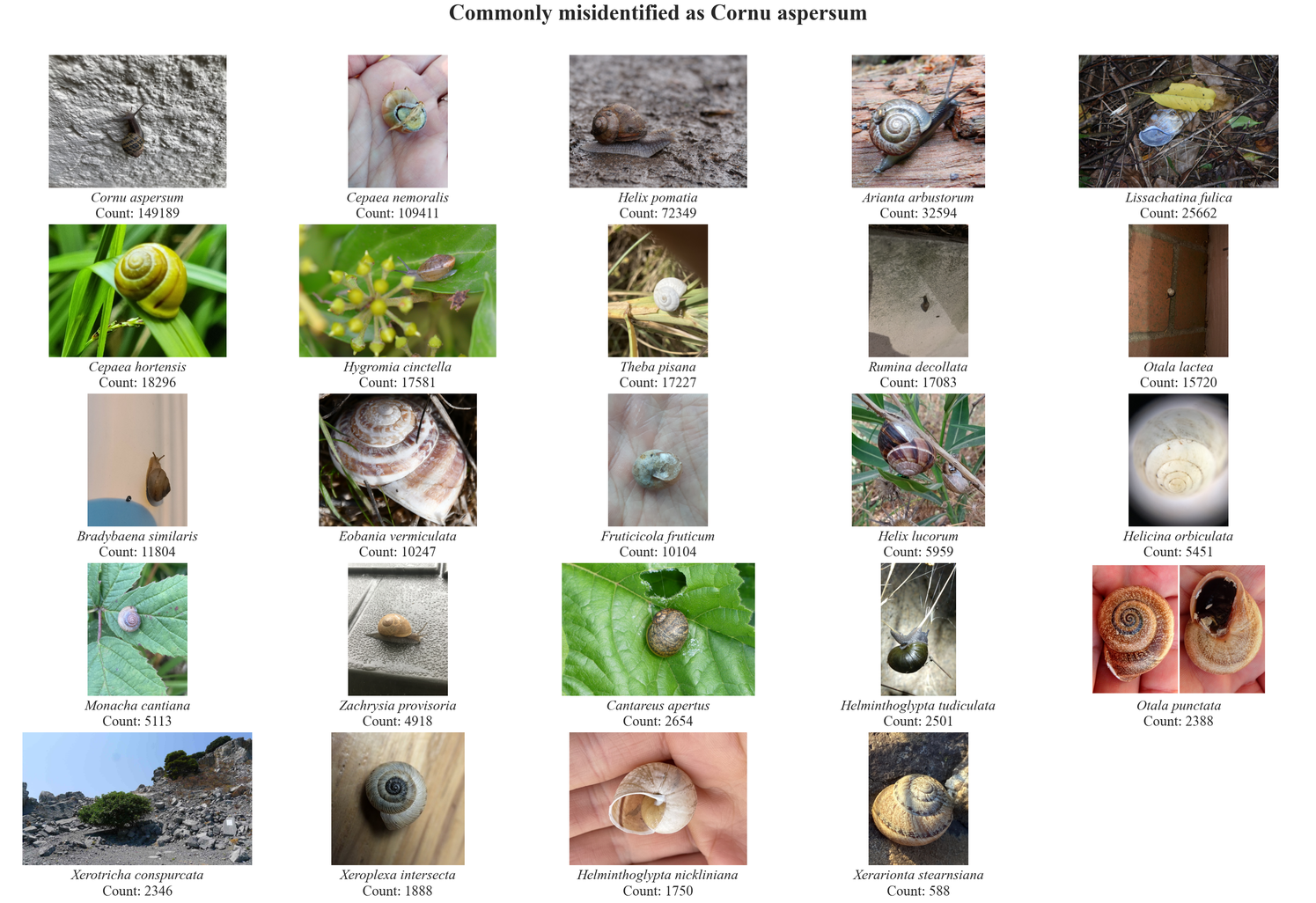}
    \caption{Sample of commonly misidentified of selected species (\textit{Mollusca}) \textit{Cornu aspersum}.}
    \label{fig:Cornu_aspersum}
\end{figure}

\begin{figure}[H]
    \centering
    \includegraphics[width=\textwidth]{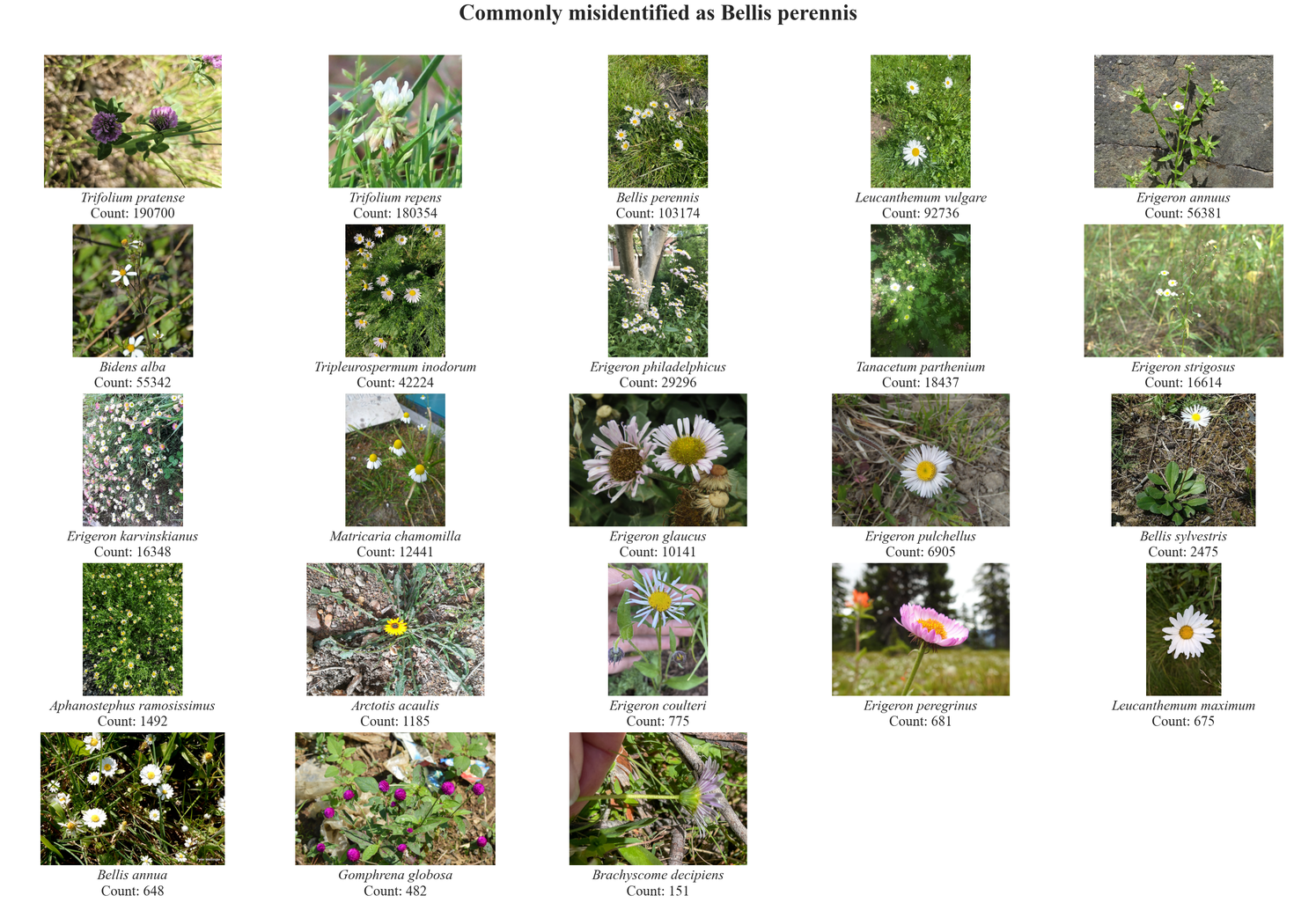}
    \caption{Sample of commonly misidentified of selected species (\textit{Plantae}) \textit{Bellis perennis}.}
    \label{fig:Bellis_perennis}
\end{figure}

\begin{figure}[H]
    \centering
    \includegraphics[width=\textwidth]{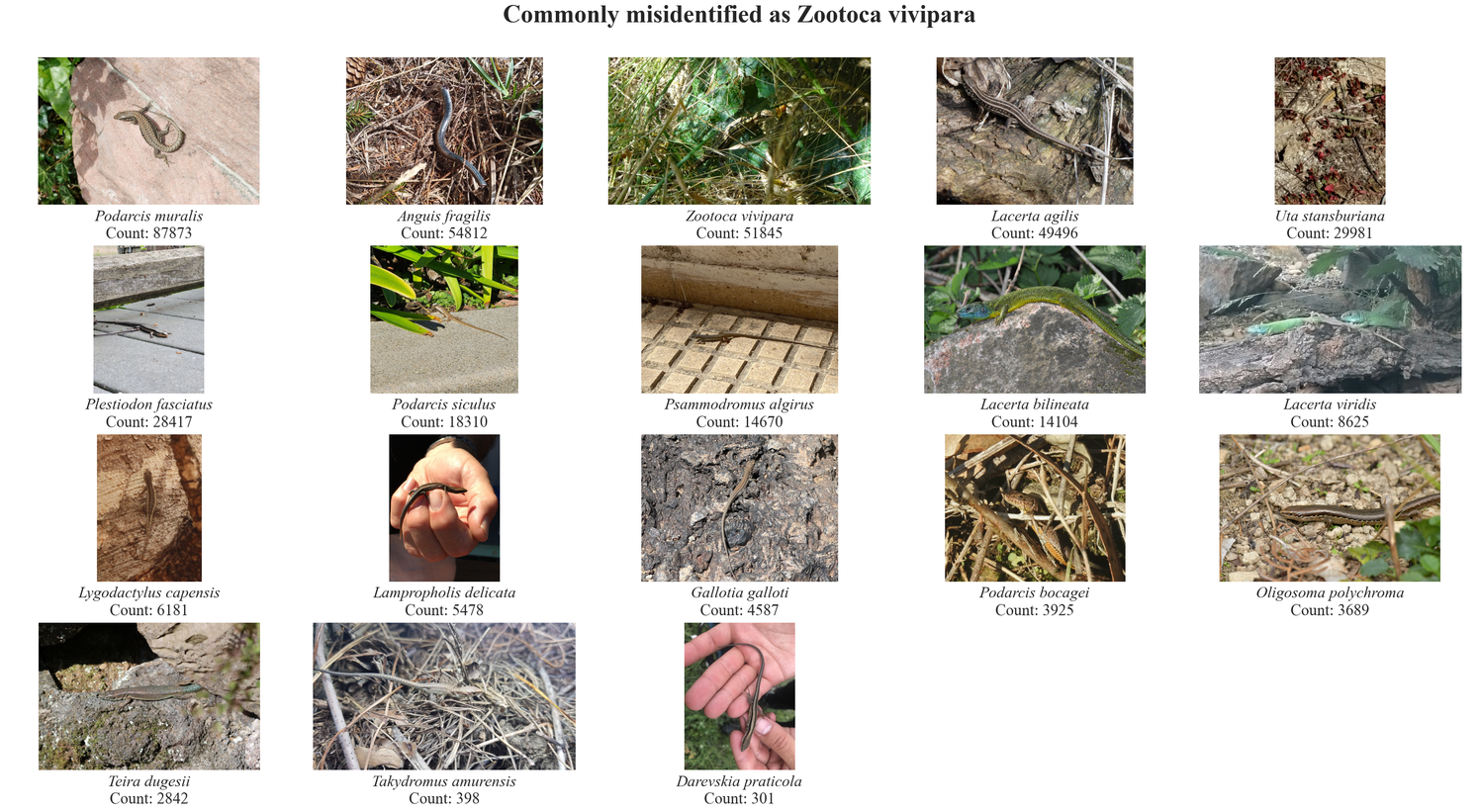}
    \caption{Sample of commonly misidentified of selected species (\textit{Reptilia}) \textit{Zootoca vivipara}.}
    \label{fig:Zootoca_vivipara}
\end{figure}

We randomly select  species from each taxonomic group \textit{Arachnida}, \textit{Aves}, \textit{Fungi},  \textit{Insecta}, \textit{Mollusca}, \textit{Plantae}, and \textit{Reptilia} and corresponding visually confusion group species for each benchmark. Figure~\ref{fig:Parasteatoda_tepidariorum}--\ref{fig:Zootoca_vivipara} show examples of cryptic groups in \textsc{CrypticBio}. For benchmarking we randomly select 100 images for each species.


\subsection{\textsc{CrypticBio-CommonUnseen} benchmark details}
We strictly select \textsc{CrypticBio-Common} to taxa observed from 01-09-2024 to 01-04-2025. This, we ensure that we evaluate zero-shot learning of established state-of-the-art models using new observations (i.e. images). We randomly select 100 images for each species, spanning n = 133 species (26 species less than \textsc{CrypticBio-Common}).


\subsection{\textsc{CrypticBio-Endangered} benchmark details}
To highlight the challenges of species identification within conservation-critical contexts, we introduce \textsc{CrypticBio-Endangered}, a curated subset of cryptic species that are listed as threatened or endangered according to the global IUCN Red List~\cite{iucn2024}. This subset is designed to assess model performance on taxa where misidentification may carry heightened ecological and conservation risks.

We select a species from each  taxonomic groups—\textit{Arachnida},  \textit{Aves}, \textit{Fungi}, \textit{Insecta}, \textit{Mollusca}, \textit{Plantae}, and \textit{Reptilia}—each of which contains species that are both visually similar and conservation-relevant. From each group, we randomly select 100 representative samples and their corresponding cryptic species groups. To ensure data quality and sufficient representation for evaluation, we filter out taxa with fewer than 150 recorded observations.

This subset emphasizes the importance of accurate classification for threatened taxa, where even minor identification errors can undermine conservation priorities and downstream ecological analyses. Table \ref{table:endangered_stats} and Figure further detail the sample characteristics.

\begin{table}[H]
\begin{center}
\small
\caption{\textsc{CrypticBio-Endangered} subset distribution.}
\label{table:endangered_stats}
\begin{tabular}{c c c c} 
\toprule
\textbf{Taxon} & \textbf{Selected species} & \textbf{\#Associated cryptic species} & \textbf{\#Observations}\\
\midrule
(\textit{Arachnida}) & \textit{Dolomedes plantarius} & 2 & 2352\\
(\textit{Aves}) & \textit{Calidris ruficollis} & 4 & 129885 \\
(\textit{Fungi}) & \textit{Hygrocybe intermedia} & 3 & 16540\\ 
(\textit{Insecta}) & \textit{Petalura gigantea} & 3 & 739\\ 
(\textit{Mollusca}) & \textit{Pinna nobilis} & 4 & 2592\\
(\textit{Plantae}) & \textit{Guaiacum officinale} & 6 &  13981\\
(\textit{Reptilia}) & \textit{Vipera aspis vivipara} & 15 & 126034\\
\bottomrule
\end{tabular}
\end{center}
\end{table}

\subsection{\textsc{CrypticBio-Invasive} benchmark details}
To address the increasing ecological risks posed by invasive alien species (IAS), we introduce \textsc{CrypticBio-Invasive}, a dedicated benchmark subset focusing on invasive species and their cryptic species selected from the 100 of the World's Worst Invasive Alien Species by Global Invasive Species Database (GISD)~\cite{gisd2025}. 
IAS are recognized as a major driver of biodiversity loss, with their occurrences showing exponential growth worldwide~\cite{mormul2022invasive}. 
Accurate identification of invasive taxa—particularly those embedded within morphologically cryptic species complexes—is therefore critical for early detection, monitoring, and mitigation efforts.

In constructing this subset, we select 100 representative samples from each cryptic group associated with a selected invasive species. 
To ensure statistical robustness and adequate representation, we exclude taxa for which fewer than 150 validated observations are available. We select one species for each taxons (i.e. \textit{Aves}, \textit{Fungi}, \textit{Insecta}, \textit{Mollusca}, \textit{Plantae}, excluding \textit{Arachnida} and \textit{Reptilia} as there are not species mentionings in 100 of the World's Worst Invasive Alien Species by Global Invasive Species Database (GISD)). 

\textsc{CrypticBio-Invasive} highlights the unique challenge of identifying invasive taxa that are visually indistinguishable from native or non-invasive relatives, and serves as a targeted testbed for evaluating model performance in scenarios with direct ecological and policy implications.
Table \ref{table:invasive_stats} further detail the sample characteristics.

\begin{table}[t]
\begin{center}
\small
\caption{\textsc{CrypticBio-Invasive} subset distribution.}
\label{table:invasive_stats}
\begin{tabular}{c c c c} 
\toprule
\textbf{Taxonomic group} & \textbf{Selected species} & \textbf{\#Associated cryptic group} & \textbf{\#Observations}\\
\midrule
(\textit{Aves}) & \textit{Acridotheres tristis} & 25 & 34689\\
(\textit{Fungi}) & \textit{Cryphonectria parasitica} & 4 & 672\\
(\textit{Insecta}) & \textit{Linepithema humile} & 24 & 15178\\ 
(\textit{Plantae}) &\textit{Acacia mearnsii} & 25 & 14579 \\
\bottomrule
\end{tabular}
\end{center}
\end{table}

\section{\textsc{Extended results}}
\label{supplementary_material:extended_results}
\textbf{Experimental details} We benchmark state-of-the-art CLIP-style biodiversity models and the species taxonomic level. We evaluate on Nvidia GeForce RTX 3080 GPU using 32 GB of RAM memory. Table \ref{table:model_suite} summaries our model suite.
We use \textsc{BioCLIP} \cite{stevens2024bioclip}, \textsc{BioTrove}'s \textsc{BioCLIP} ViT-B-16 and OpenAI ViT-B-16 fine-tuned variants \cite{yang2024biotrove}, and \textsc{TaxaBind} \cite{sastry2024taxabind} as image-only baseline models.

For multimodal evaluation, we add embeddings obtained from the image encoders to those obtained from \textsc{TaxaBind} location and environmental features encoders, which are then used for zero-shot classification. We collect from WorldClim-2.1\cite{fick2017worldclim} environmental features for each observation based on the location metadata,
which are then passed through \textsc{TaxaBind}'s environmental encoder. We compare performance on scientific, vernacular, and mixed text types, as advised in \cite{stevens2024bioclip}. We combine image with location and environment embeddings by adding each embedding. 

We benchmark on all English available vernacular terminology, and we use species scientific name when vernacular term is missing. 
Tables \ref{table:extended_results_common}--\ref{table:extended_results_invasive} summarize performance comparisons across benchmarks on \textsc{CrypticBio-Common}, \textsc{CrypticBio-CommonUnseen}, \textsc{CrypticBio-Endangered}, and \textsc{CrypticBio-Common-Invasive}.

We report zero-shot top-1 accuracy based on cosine similarity.
We include a 95\% confidence intervals for all reported metrics, calculated using binomial proportion confidence interval method (denoted as $\pm$). 
Furthermore, we compute an aggregate performance metric, which represents the weighted average accuracy over all classes across the benchmarks.
To assess the significance of pairwise performance differences between models, we use McNemar’s test (p-value < 0.05).

\begin{table}[t]
\centering
\small
\caption{Model suite used in our benchmark.}
\label{table:model_suite}
\begin{tabular}{c l l c} 
\toprule
\textbf{Modality} &\textbf{Model} & & \textbf{Architecture} \\
\midrule
\multirow{4}{*}{
Image}& \textbf{\textsc{BioCLIP}} \cite{stevens2024bioclip} & \textbf{BC } & ViT-B-16 \\
\cmidrule(r){2-4}
& \textbf{\textsc{BioTrove} \textsc{BioCLIP}} \cite{yang2024biotrove} & \textbf{BT-B} & ViT-B-16\\
& \textbf{\textsc{BioTrove} \textsc{OpenAI}} \cite{yang2024biotrove} & \textbf{BT-O} & ViT-B-16\\
\cmidrule(r){2-4}
& \textbf{\textsc{TaxaBind}} \cite{sastry2024taxabind} & \textbf{TB} & ViT-B-16\\
\midrule
Geographical location & \textbf{\textsc{TaxaBind}} \cite{sastry2024taxabind} & \textbf{L} & \textsc{GeoCLIP} \cite{vivanco2023geoclip} \\
\midrule
Environment features & \textbf{\textsc{TaxaBind}} \cite{sastry2024taxabind} & \textbf{E} & ResNet-Style MLP \cite{cole23} \\
\bottomrule
\end{tabular}
\end{table}

\begin{table}[t]
\centering
\small
\caption{Example of benchmarked text types.}
\label{table:text_types}
\begin{tabular}{l l} 
\toprule
\textbf{Text type} & \textbf{Example} \\
\midrule
Scientific & \textit{Bellis perennis}\\
Vernacular & \textit{Common daisy}, \textit{English daisy}, \textit{Lawn daisy}\\
Scientific + Vernacular & \textit{Bellis perennis} commonly known as \textit{Common daisy}, \textit{English daisy}, \textit{Lawn daisy}\\
\bottomrule
\end{tabular}
\end{table}

\textbf{Results overview} 
We find combining image and location embeddings to improve performance on zero-shot image classification overall. Models trained with specialist datasets (i.e. \textsc{BioTrove-CLIP} and \textsc{BioCLIP}) perform better. Additionally, mixed scientific and common names yield overall best performance scores, thus, we only report these scores. 
 
We additionally evaluate \textsc{CrypticBio-CommonUnseen} across all \textsc{BioTrove} variants, noting that this set comprises taxa observations entirely held out from training.

\begin{table}[H]
\centering
\caption{\textsc{CrypticBio-CommonUnseen} benchmark on various models. I / L / E refers to image / location / environmental features embeddings; AR / AV / F / I / M / P / R refers to taxonomic groups \textit{Arachnida} / \textit{Aves} / \textit{Fungi} / \textit{Insecta} / \textit{Mollusca} / \textit{Plantae} / \textit{Reptilia}; MN refers to mixed (scientific + common) text annotations; WA refers to weighted average; BC refers to \textsc{BioCLIP}; BT-B refers to \textsc{BioTrove-CLIP-BioCLIP}; BT-O refers to \textsc{BioTrove-CLIP-OpenCLIP}; TB refers to \textsc{TaxaBind}. Location (L) and environmental features (E) are \textsc{TaxaBind} embeddings.}
\label{table:extended_results_common}
\scriptsize
\begin{tabular}{l c c c c c c c c c} 
\toprule
\textbf{Model} & \textbf{Modality} & \textbf{AR-MN} & \textbf{AV-MN} & \textbf{F-MN} & \textbf{I-MN} & \textbf{M-MN} & \textbf{P-MN} & \textbf{R-MN} & \textbf{WA} \\
\midrule
\textbf{BC} & I & 37.6 {\scriptsize ±1.58} & 55.1 {\scriptsize ±1.62} & 49.7 {\scriptsize ±1.63} & 38.5 {\scriptsize ±1.59} & 29.5 {\scriptsize ±1.49} & 63.7 {\scriptsize ±1.57} & 46.0 {\scriptsize ±1.63} & 45.76 \\
\textbf{BT-B} & I & 59.1 {\scriptsize ±1.61} & 61.4 {\scriptsize ±1.59} & 72.7 {\scriptsize ±1.45} & 50.5 {\scriptsize ±1.63} & 49.2 {\scriptsize ±1.63} & \textbf{76.4 {\scriptsize ±1.39}} & 62.0 {\scriptsize ±1.58} & 61.66 \\
\textbf{BT-O} & I & 50.8 {\scriptsize ±1.63} & 43.4 {\scriptsize ±1.62} & 74.5 {\scriptsize ±1.42} & 31.7 {\scriptsize ±1.52} & 39.7 {\scriptsize ±1.60} & 60.0 {\scriptsize ±1.60} & 48.4 {\scriptsize ±1.63} & 49.84 \\
\textbf{TB} & I & 41.2 {\scriptsize ±1.61} & 59.5 {\scriptsize ±1.60} & 52.5 {\scriptsize ±1.63} & 40.4 {\scriptsize ±1.60} & 32.4 {\scriptsize ±1.53} & 64.0 {\scriptsize ±1.57} & 52.0 {\scriptsize ±1.63} & 48.89 \\
\midrule
\textbf{TB} & I+L & 41.4 {\scriptsize ±1.61} & 59.6 {\scriptsize ±1.60} & 52.6 {\scriptsize ±1.63} & 40.5 {\scriptsize ±1.60} & 32.5 {\scriptsize ±1.53} & 64.2 {\scriptsize ±1.56} & 52.5 {\scriptsize ±1.63} & 49.08 \\
\textbf{BT-B} & I+L & \textbf{59.5 {\scriptsize ±1.60}} & \textbf{65.1 {\scriptsize ±1.56}} & 76.3 {\scriptsize ±1.39} & \textbf{54.9 {\scriptsize ±1.62}} & \textbf{50.2 {\scriptsize ±1.63}} & 75.0 {\scriptsize ±1.41} & \textbf{68.7 {\scriptsize ±1.51}} & \textbf{64.29} \\
\textbf{BT-B} & I+E & 25.2 {\scriptsize ±1.42} & 30.5 {\scriptsize ±1.50} & 37.5 {\scriptsize ±1.58} & 14.0 {\scriptsize ±1.14} & 21.5 {\scriptsize ±1.34} & 26.8 {\scriptsize ±1.45} & 27.2 {\scriptsize ±1.45} & 26.14 \\
\textbf{BT-O} & I+L & 50.4 {\scriptsize ±1.63} & 46.0 {\scriptsize ±1.63} & \textbf{76.4 {\scriptsize ±1.39}} & 34.4 {\scriptsize ±1.55} & 38.7 {\scriptsize ±1.59} & 63.0 {\scriptsize ±1.58} & 53.5 {\scriptsize ±1.63} & 51.79 \\
\textbf{BT-O} & I+E & 23.1 {\scriptsize ±1.38} & 21.9 {\scriptsize ±1.35} & 40.7 {\scriptsize ±1.60} & 12.4 {\scriptsize ±1.08} & 16.4 {\scriptsize ±1.21} & 20.4 {\scriptsize ±1.32} & 21.1 {\scriptsize ±1.33} & 22.33 \\
\bottomrule
\end{tabular}
\end{table}

\begin{table}[H]
\centering
\caption{\textsc{CrypticBio-Endangered} benchmark on various models. I / L / E refers to image / location / environmental features embeddings; AR / AV / F / I / M / P / R refers to taxonomic groups \textit{Arachnida} / \textit{Aves} / \textit{Fungi} / \textit{Insecta} / \textit{Mollusca} / \textit{Plantae} / \textit{Reptilia}; MN refers to mixed (scientific + common) text annotations; WA refers to weighted average; BC refers to \textsc{BioCLIP}; BT-B refers to \textsc{BioTrove-CLIP-BioCLIP}; BT-O refers to \textsc{BioTrove-CLIP-OpenCLIP}; TB refers to \textsc{TaxaBind}. Location (L) and environmental features (E) are \textsc{TaxaBind} embeddings.}
\label{table:extended_results_endangered}
\scriptsize
\begin{tabular}{l c c c c c c c c c} 
\toprule
\textbf{Model} & \textbf{Modality} & \textbf{AR-MN} & \textbf{AV-MN} & \textbf{F-MN} & \textbf{I-MN} & \textbf{M-MN} & \textbf{P-MN} & \textbf{R-MN} & \textbf{WA} \\
\midrule
\textbf{BC} & I & 53.0 {\scriptsize ±1.63} & 49.5 {\scriptsize ±1.63} & 60.3 {\scriptsize ±1.60} & 74.3 {\scriptsize ±1.43} & 33.0 {\scriptsize ±1.54} & 60.3 {\scriptsize ±1.60} & 27.3 {\scriptsize ±1.45} & 51.11 \\
\textbf{BT-B} & I & 48.0 {\scriptsize ±1.63} & 43.8 {\scriptsize ±1.62} & 50.7 {\scriptsize ±1.63} & 48.3 {\scriptsize ±1.63} & \textbf{44.8 {\scriptsize ±1.62}} & 51.3 {\scriptsize ±1.63} & \textbf{34.4 {\scriptsize ±1.55}} & 45.89 \\
\textbf{BT-O} & I & 45.0 {\scriptsize ±1.62} & 26.8 {\scriptsize ±1.45} & 64.7 {\scriptsize ±1.56} & 46.7 {\scriptsize ±1.63} & 36.8 {\scriptsize ±1.57} & 47.0 {\scriptsize ±1.63} & 21.2 {\scriptsize ±1.33} & 41.15 \\
\textbf{TB} & I & \textbf{54.0 {\scriptsize ±1.63}} & \textbf{53.3 {\scriptsize ±1.63}} & 59.7 {\scriptsize ±1.60} & \textbf{75.0 {\scriptsize ±1.41}} & 31.8 {\scriptsize ±1.52} & \textbf{63.2 {\scriptsize ±1.57}} & 31.5 {\scriptsize ±1.52} & \textbf{52.62} \\
\midrule
\textbf{TB} & I+L & \textbf{54.0 {\scriptsize ±1.63}} & 53.0 {\scriptsize ±1.63} & 59.3 {\scriptsize ±1.60} & 74.7 {\scriptsize ±1.42} & 31.5 {\scriptsize ±1.52} & \textbf{63.3 {\scriptsize ±1.57}} & 31.6 {\scriptsize ±1.52} & 52.48 \\
\textbf{BT-B} & I+L & 46.5 {\scriptsize ±1.63} & 47.8 {\scriptsize ±1.63} & 45.0 {\scriptsize ±1.62} & 44.7 {\scriptsize ±1.62} & 44.5 {\scriptsize ±1.62} & 55.5 {\scriptsize ±1.62} & 32.8 {\scriptsize ±1.53} & 45.24 \\
\textbf{BT-B} & I+E & 43.5 {\scriptsize ±1.62} & 37.0 {\scriptsize ±1.58} & 33.3 {\scriptsize ±1.54} & 33.3 {\scriptsize ±1.54} & 31.0 {\scriptsize ±1.51} & 33.5 {\scriptsize ±1.54} & 19.4 {\scriptsize ±1.29} & 33.01 \\
\textbf{BT-O} & I+L & 42.0 {\scriptsize ±1.61} & 28.0 {\scriptsize ±1.47} & \textbf{65.0 {\scriptsize ±1.56}} & 40.0 {\scriptsize ±1.60} & 42.0 {\scriptsize ±1.61} & 48.2 {\scriptsize ±1.63} & 20.5 {\scriptsize ±1.32} & 40.81 \\
\textbf{BT-O} & I+E & 40.0 {\scriptsize ±1.60} & 26.0 {\scriptsize ±1.43} & 35.7 {\scriptsize ±1.56} & 33.3 {\scriptsize ±1.54} & 31.0 {\scriptsize ±1.51} & 32.8 {\scriptsize ±1.53} & 11.8 {\scriptsize ±1.06} & 30.09 \\
\bottomrule
\end{tabular}
\end{table}

\begin{table}[H]
\centering
\caption{\textsc{CrypticBio-Invasive} benchmark on various models. I / L / E refers to image / location / environmental features embeddings; AV / F / I / P refers to taxonomic groups \textit{Aves} / \textit{Fungi} / \textit{Insecta} / \textit{Plantae}; MN refers to mixed (scientific + common) text annotations; WA refers to weighted average; BC refers to \textsc{BioCLIP}; BT-B refers to \textsc{BioTrove-CLIP-BioCLIP}; BT-O refers to \textsc{BioTrove-CLIP-OpenCLIP}; TB refers to \textsc{TaxaBind}. Location (L) and environmental features (E) are \textsc{TaxaBind} embeddings.}
\label{table:extended_results_invasive}
\scriptsize
\begin{tabular}{l c c c c c c} 
\toprule
\textbf{Model} & \textbf{Modality} & \textbf{AV-MN} & \textbf{F-MN} & \textbf{I-MN} & \textbf{P-MN} & \textbf{WA} \\
\midrule
\textbf{BC} & I & 59.91 {\scriptsize ±1.60} & 66.75 {\scriptsize ±1.54} & 33.61 {\scriptsize ±1.54} & 36.17 {\scriptsize ±1.57} & 49.11 \\
\textbf{BT-B} & I & 76.05 {\scriptsize ±1.39} & 61.25 {\scriptsize ±1.59} & 54.72 {\scriptsize ±1.63} & 41.42 {\scriptsize ±1.61} & 58.36 \\
\textbf{BT-O} & I & 62.15 {\scriptsize ±1.58} & 55.75 {\scriptsize ±1.62} & 48.11 {\scriptsize ±1.63} & 29.77 {\scriptsize ±1.49} & 48.95 \\
\textbf{TB} & I & 64.12 {\scriptsize ±1.57} & \textbf{69.00 {\scriptsize ±1.51}} & 38.28 {\scriptsize ±1.59} & 37.58 {\scriptsize ±1.58} & 52.24 \\
\midrule
\textbf{TB} & I+L & 64.32 {\scriptsize ±1.56} & 68.75 {\scriptsize ±1.51} & 38.61 {\scriptsize ±1.59} & 37.58 {\scriptsize ±1.58} & 52.31 \\
\textbf{BT-B} & I+L & \textbf{76.77 {\scriptsize ±1.38}} & 64.75 {\scriptsize ±1.56} & \textbf{61.06 {\scriptsize ±1.59}} & \textbf{50.33 {\scriptsize ±1.63}} & \textbf{63.23} \\
\textbf{BT-B} & I+E & 18.78 {\scriptsize ±1.28} & 57.50 {\scriptsize ±1.61} & 21.39 {\scriptsize ±1.34} & 22.99 {\scriptsize ±1.37} & 30.17 \\
\textbf{BT-O} & I+L & 61.79 {\scriptsize ±1.59} & 57.25 {\scriptsize ±1.62} & 50.06 {\scriptsize ±1.63} & 33.03 {\scriptsize ±1.54} & 50.53 \\
\textbf{BT-O} & I+E & 13.18 {\scriptsize ±1.10} & 52.75 {\scriptsize ±1.63} & 18.11 {\scriptsize ±1.26} & 14.21 {\scriptsize ±1.14} & 24.56 \\
\bottomrule
\end{tabular}
\end{table}

We deliberately evaluate our approach in a zero-shot setting to assess its generalization capabilities without relying on task-specific fine-tuning. This is an intentional choice aimed at evaluating how well additional contextual information can be leveraged within already existing biodiversity models.


\end{document}